%% file: JournalRevision.tex
\documentclass[11pt]{article}
\input{packages.tex}

\input{standard.tex}

\input{AlgMacros.tex}

\usepackage{float}

\newcommand{\thistitle}{Crossed Products, Conditional Expectations\\ and Constraint Quantization}

\begin{document}

\title{\thistitle}
\author{
	Marc S. Klinger 
	and 
	Robert G. Leigh 
	\\
	\\
	{\small \emph{\uiuc{a}}}
	\\
	}
\date{}
\maketitle
\vspace{-0.5cm}
\begin{abstract}
Recent work has highlighted the importance of crossed products in correctly elucidating the operator algebraic approach to quantum field theories. In the gravitational context, the crossed product simultaneously promotes von Neumann algebras associated with subregions in diffeomorphism covariant quantum field theories from type III to type II, and provides the necessary ingredients to gravitationally dress operators, thereby enforcing the constraints of the theory. In this note we enhance the crossed product construction to the context of general gauge theories with arbitrary combinations of internal and spacetime local symmetries. This is done by leveraging the correspondence between the crossed product and the extended phase space. We then undertake a detailed study of constraint quantization from the perspective of a generic crossed product algebra. We study and compare four distinct approaches to constraint quantization from this point of view: refined algebraic quantization, BRST quantization, path integral quantization, and the commutation theorem for crossed products. Far from simply reproducing existing analyses, the operator algebraic viewpoint sheds new light on old problems by reformulating the dressing of operators in terms of conditional expectations and other closely related projection maps. We conclude by applying our approach to the constraint quantization of three distinct gauge theories including a discussion of gravity on null hypersurfaces.
\vspace{0.3cm}
\end{abstract}

\pagebreak

\begingroup
\hypersetup{linkcolor=black}
\tableofcontents
\endgroup
\noindent\rule{\textwidth}{0.6pt}

\setcounter{footnote}{0}
\renewcommand{\thefootnote}{\arabic{footnote}}

\newcommand{\curr}[1]{\mathbb{J}_{#1}}
\newcommand{\constr}[1]{\mathbb{M}_{#1}}
\newcommand{\chgdens}[1]{\mathbb{Q}_{#1}}
\newcommand{\spac}[1]{S_{#1}}
\newcommand{\hyper}[1]{\Sigma_{#1}}
\newcommand{\chg}[1]{\mathbb{H}_{#1}}
\newcommand{\ThomSig}[1]{\hat{\Sigma}_{#1}}
\newcommand{\ThomS}[1]{\hat{S}_{#1}}
\newcommand{\discuss}[1]{{\color{red} #1}}

\newcommand{\gravgroup}{G_L}
\newcommand{\gaugegroup}{G_G}
\newcommand{\gravL}{L_L}
\newcommand{\gaugeL}{L_G} 
\newcommand{\solder}{e}
\newcommand{\sym}{\Omega}
\newcommand{\symdens}{\omega}
\newcommand{\sympot}{\Theta}
\newcommand{\sympotdens}{\theta}
\newcommand{\chgcon}[1]{\chg{#1}^{(1)}}
\newcommand{\chgchg}[1]{\chg{#1}^{(2)}}
\newcommand{\cpalg}{\mathcal{M}_{cp}}

\newcommand{\red}[1]{{\color{red} #1}}
\newcommand{\purple}[1]{{\color{purple} #1}}
\pagebreak

\section{Introduction}

In recent work \cite{Klinger:2023tgi} we have argued that the operator algebra associated with a subregion in a diffeomorphism covariant quantum field theory is more well-behaved if one includes degrees of freedom associated with diffeomorphisms of the  interface between the given subregion and its complement. By more well-behaved we mean  that the von Neumann algebra associated with quantum fields in the subregion goes from type III to type II. This implies that, in contrast to the naive subregion algebra, the enhanced algebra possesses a faithful semi-finite trace, admits density operators, and allows for a meaningful notion of {entanglement entropy} to be assigned to states in the theory (up to an additive constant). The algebraic mechanism underlying the improvement to the subregion theory is the implementation of a crossed product between the naive subregion algebra and its modular automorphism group \cite{takesaki1973crossed,Connes1973, takesaki2003theory}. The relationship between the modular automorphism group and the diffeomorphism covariance of the theory is explicated in \cite{Klinger:2023tgi} by appealing to the correspondence between modular Hamiltonians for subregion quantum field theories and diffeomorphisms of entanglement cuts therein \cite{faulkner2016shape, faulkner2016modular}. 

An alternative approach to realizing the algebra appropriate to a subregion in the presence of semi-classical gravity has been undertaken in \cite{Witten:2021unn, Chandrasekaran:2022eqq, Chandrasekaran:2022cip,Jensen:2023yxy} building on \cite{Leutheusser:2021qhd,Leutheusser:2021frk,Leutheusser:2022bgi} in which large $N$ von Neumann algebras were studied in the context of the AdS/CFT correspondence.\footnote{Many other recent works have built on the role of the crossed product, these include: \cite{AliAhmad:2023etg,Ali:2023bmm, Soni:2023fke, Bahiru:2023ify, Kudler-Flam:2023qfl}.} In these papers the crossed product algebra is motivated by introducing an ``observer" who provides a point of reference through which observables of the theory can be ``gravitationally dressed". By dressing one means a procedure that alters observables so as to ensure that they commute with a set of operators that represent constraints of the theory; in the gravitational context these constraints correspond to the imposition of diffeomorphism invariance. The role of the observer to this end is quite clear -- mandating that operators always be defined with respect to the worldline of the observer turns the measurement of such operators into a relative object. Qualitatively, one may therefore regard the observer as defining a ``clock" relative to which meaningful experiments can be conducted. From this point of view the crossed product is instrumental in accomplishing two ends -- firstly, as was the case above, it enhances the algebra from type III to type II by incorporating the generators of the modular automorphism, and secondly it provides the necessary ingredients for implementing the constraints of the theory. 

The introduction of an observer may be regarded as a resolution of the dilemma that a theory is not complete if it does not furnish the tools necessary to distinguish physical from unphysical degrees of freedom. Put differently, a theory is made meaningful only once it is equipped with a list of items which it can be used to measure. With that being said, the resolution offered by the observer to this problem is incomplete in a different way. Its most glaring defect, as acknowledged by the authors, is that the observer is introduced in a fashion that is distinctly separate from the rest of the relevant features of the theory. In a fully satisfactory theory both the system that is being measured and the tools being used to make measurements should be made up of the same ``stuff". A desire to rectify this problem was largely the motivation behind \cite{Klinger:2023tgi}. A central result from that note is that the same crossed product algebra that is implicated by the introduction of the observer can be realized by carefully ensuring that all the symmetries of a given system are equivariantly realized in an appropriate prequantum phase space.

The aforementioned fact is a specific example of a more general correspondence between the crossed product of a von Neumann algebra with a group of automorphisms, and the extension of a symplectic phase space to account for a group of automorphisms acting on its Poisson algebra. We refer to the latter construction as the extended phase space \cite{Ciambelli:2021nmv,Klinger:2023qna}. The formal correspondence between the crossed product and the extended phase space is summarized in the commutative diagram: 
\begin{equation} \label{commutative diagram}
\begin{tikzcd}
(X,\Omega) \arrow[r, "Q"] \arrow[d, "EPS"]
& \mathcal{M} \arrow[d, "CP"] \\
(X_{ext},\Omega_{ext}) \arrow[r, "Q"]
& \mathcal{M} \rtimes G
\end{tikzcd}
\end{equation} 
Here, $EPS$ and $CP$ are, respectively, the extension of a phase space by the group\footnote{As will be stressed later on in the note the symmetry structures that appear in gauge theories are often Lie groupoids as opposed to Lie groups. To avoid overloading language we will often simply use the word group. However, there is important content in passing from a group to a groupoid, especially for diffeomorphism invariant theories, see Appendix E of \cite{Klinger:2023tgi}.} $G$ and the crossed product of an algebra by the group $G$, and $Q$ is a valid quantization procedure which takes as input the data of a symplectic manifold and outputs an algebra of quantum observables. From the point of view of the extended phase space, the notion of an observer emerges automatically  as a consequence of the fact that the (gauge) symmetries are represented equivariantly. We will explain this structure formally as well as through a series of examples later in the paper.  

In this note we aim to extend this previous work in two specific directions. First, we will demonstrate how the correspondence \eqref{commutative diagram} allows for the formulation of a complete algebra of operators in arbitrary quantum gauge theories. Secondly, we undertake a detailed study of constraint quantization from the operator algebraic perspective highlighting the central role of the extended phase space and its associated crossed product algebra in guiding and unifying various approaches to implementing constraints quantum mechanically. As mentioned, our previous work largely emphasized the role of the crossed product in enhancing subregion algebras from type III to type II as part of an effort to understand how geometry naturally regularizes entanglement entropy. We did not provide there an equally complete discussion of how the crossed product, when constructed via the correspondence \eqref{commutative diagram} rather than through the inclusion of an observer, allows for the implementation of constraints in the vein of \cite{Witten:2021unn, Chandrasekaran:2022eqq, Chandrasekaran:2022cip,Jensen:2023yxy}. Thus, in this note we concentrate our attention on this facet of the crossed product. It is worth mentioning that, in recent work by one of the authors \cite{Ciambelli:2023mir}, the symplectic analysis of the Raychaudhuri equation on null hypersurfaces was given in the context of an arbitrary gravity theory coupled to matter. One of the main insights of that work is an explicit realization of the role of extended degrees of freedom in defining observer like variables that allow for the satisfaction of constraints. We will comment further on this example later in the paper.

The paper is organized into two major technical sections, along with a section that presents several examples. In Section \ref{sec: preliminaries} we review the geometric approach to gauge theories in terms of Atiyah Lie algebroids, construct their prequantum phase space, and describe their subsequent quantization (albeit only qualitatively).\footnote{In forthcoming work we will address the problem of quantization more directly. For the purpose of this note, we assume the existence of a quantization functor lifting symplectic data into an operator algebra.} The first major advantage of formulating the theory in this way is that it provides a framework in which both internal symmetries and diffeomorphisms are treated on an equal footing as gauge symmetries, and, moreover, in which these gauge symmetries are all faithfully realized in the extended phase space of the theory. A different way of phrasing the previous result is that the extended phase space of the gauge theory constructs an equivariant moment map into its Poisson algebra from a Lie algebroid whose elements correspond to generically field-dependent spacetime local gauge transformations and diffeomorphisms. The existence of such a moment map is a critical feature because it implies that approaches such as geometric quantization can, in principle, be applied to the quantization of the (unconstrained) gauge theory. 

The second major advantage of this approach is that it allows for a distinction to be made between symmetries which are pure redundancies of a given theory and symmetries which are physically realized in the sense that they map physically distinct states to physically distinct states. In this regard, we can treat the extended phase space of a gauge theory as possessing two copies of the group of gauge transformations and diffeomorphisms: one which furnishes a set of constraints that will need to be enforced and one which merely introduces an additional set of physical charges, a structure that is implied by Noether's second theorem. As has been a major theme in recent work \cite{Ciambelli:2021vnn,Ciambelli:2022vot,Ciambelli:2022cfr,Donnelly:2016auv,Freidel:2020xyx,Freidel:2020svx,Freidel:2020ayo,Freidel:2021cjp,Freidel:2023bnj,Speranza:2017gxd,Donnelly:2022kfs,Speranza:2022lxr,donnelly2021gravitational}, the physical charges in a gauge theory are intimately related with the presence of codimension two submanifolds in spacetime which are often referred to as  \emph{corners}. 

After having given an overview of the prequantum phase space associated with a general gauge theory, we employ the correspondence introduced in \cite{Klinger:2023tgi} to identify the quantized algebra of operators as the result of a pair of crossed products which realize unitary operators associated with both constraints and physical charges. We refer to these extensions respectively as the extension at codimension one and the extension at codimension two. Organizing the extended phase space in terms of this pair of group automorphisms sheds some light on the multifaceted role of the crossed product previously mentioned. The extension at codimension two, which was explored in depth in \cite{Klinger:2023tgi}, ensures that corner supported charges are realized as Hamiltonian functions in the prequantum theory and quantized as physical operators in the quantum theory. An example of such an operator is the modular Hamiltonian associated with corner supported diffeomorphisms. Thus, the extension at codimension two is responsible for ensuring that the quantized algebra is always semi-finite. By contrast, the extension at codimension one furnishes a unitary representation of the constraint group which is critical for the implementation of constraints in the quantum theory. 

This brings us to Section \ref{sec: con-quant} which is the main technical thrust of the paper. The algebra realized by quantizing the full extended phase space has not yet been subjected to the constraints required by gauge symmetry. The problem of quantizing theories with constraints, or constraint quantization, was largely initiated by Dirac in \cite{dirac1964lectures}. In the intervening years constraint quantization has become a deep subject of research which lies at the heart of efforts to understand gauge theories and gravity. In this note, we will concentrate our attention on four apparently distinct approaches to constraint quantization. The first three of these approaches are hopefully quite familiar to a physics audience, they are: refined algebraic quantization (RAQ) \cite{Ashtekar:1995zh, Giulini:1998kf, Giulini:1998rk, Marolf:2000iq, Landsman:1993xe} discussed in Section \ref{sec: RAQ}, BRST quantization \cite{Batalin:1981jr,Becchi:1974xu,Becchi:1974md,Becchi:1975nq,Tyutin:1975qk,Henneaux:1992ig,Barnich:2000zw,Fuster:2005eg} discussed in Section \ref{sec: BRST}, and path integral quantization \cite{Faddeev:1967fc,Cordes:1994fc,Stora:1996ip,Weinberg:1996kr} discussed in Section \ref{sec: Path}. The existing literature on each of these approaches is vast, and the references we have included are obviously not exhaustive. Our unique contribution to this literature will be to investigate these techniques entirely from the point of view of a generic crossed product algebra $\mathcal{M} \rtimes G$, and to demonstrate in what sense $\mathcal{M} \rtimes G$ possesses all of the tools necessary to impose $G$ as a group of constraints in the style of any of the aforementioned procedures. This point of view is underscored in Section \ref{sec: comm thm} in which we discuss a fourth approach to constraint quantization by means of the commutation theorem \cite{digernes1975duality,Haagerup1978II,van1978continuous}. The commutation theorem is a powerful result from the theory of operator algebras which ensures that the full crossed product algebra can be regarded as an invariant algebra under a suitably defined set of constraints. To connect the commutation theorem with the previous approaches we provide an interpretation for it within the framework of  the correspondence \eqref{commutative diagram}. 

In summary, by formulating our analysis from an operator algebraic perspective we are able to shed new light on each of the aforementioned strategies, suggesting new remedies to overcome existing shortcomings and, perhaps most excitingly, allowing for a clear exposition on how all four forms of constraint quantization may be understood as facets of a common fundamental approach. The crucial new ingredient to this end is the structure of a (generalized) conditional expectation on a von Neumann algebra which projects from the crossed product down to the algebra of gauge invariant operators. In this regard, we recognize how the crossed product obtained by quantizing the extended phase space of a generic gauge theory allows for the implementation of a dressing procedure that generalizes \cite{Witten:2021unn, Chandrasekaran:2022eqq, Chandrasekaran:2022cip,Jensen:2023yxy}.

Although Section \ref{sec: preliminaries} and Section \ref{sec: con-quant} are conceptually linked, they are largely self-consistent and can each be read on their own. Recognizing that much of our discussion is quite formal, in Section \ref{sec: Zamples} we provide a series of worked examples in which our approach to constraint quantization is applied to well-known gauge theories. We hope that these examples are familiar to the reader, as our primary motivation in this section is to illustrate the pertinent features of our formalism and the extended phase space perspective. We begin in Section \ref{sec: Oscillator} with a quantum mechanics system, the symmetric $D$-dimensional harmonic oscillator and consider gauging the $SO(D)$ subgroup of its global symmetry, $SU(D)$. This is a useful example, because we can understand its quantization from either the Hilbert space perspective, the operator algebra perspective or the path integral perspective, and we discuss it carefully from the point of view of extended phase space. Since this is a quantum mechanics problem, it lacks one important feature, which is codimension-2 structure. In Section \ref{sec: Yang-Mills}, we therefore revisit Yang-Mills theory from the extended phase space perspective, and explain in detail how the conditional expectation interpretation can be understood through the familiar Fadeev-Popov procedure. What may not be familiar to all readers here is that we present the path integral directly in phase space, and we take pains to explain how the path integral effectively solves constraints while leaving gauge charges generally undetermined. In Section \ref{sec: Gravity} we discuss Einstein-Hilbert gravity on a null hypersurface using our conditional expectation oriented approach. Here our main focus is to elucidate the relationship between the symplectic analysis undertaken in \cite{Ciambelli:2023mir} and the extended phase space, a useful groundwork for future exploration of quantization. We conclude in Section \ref{sec: discussion} with a discussion which summarizes our results and suggests some directions for future work.

\section{Gauge Theory} \label{sec: preliminaries}

In this section we provide an outline of the operator algebra that is realized in a gauge quantum field theory starting from the extended phase space introduced in \cite{Ciambelli:2021nmv,Klinger:2023qna}.

In Appendix \ref{app: LA review} we provide an overview on how the field theoretic data of a gauge theory can be encoded in an Atiyah Lie algebroid, $A$. In this subsection, we turn our attention to the construction of the extended phase space associated with such a gauge theory. We use the notation $(A,\omega)$ to refer to a Lie algebroid with connection, drawing attention to the fact that the vector bundle $A$ endowed with different choices for the connection $\omega$ correspond to different Lie algebroids.\footnote{This can be seen most clearly through \eqref{Bracket on A} in which the bracket on $A$ is determined through the curvatures of $\omega$.} We denote by $\fldspalg$ the set of all Atiyah Lie algebroids based on the common vector bundle $A$. More concretely, we regard $\Phi \in \fldspalg$ as corresponding to the set of data specifying all possible configurations of fields in $\Omega^{\bullet}(A;E)$ including the connection $\omega$ and, if appropriate, a solder form $\solder$. 

Next, we introduce the set $\morphalg$ consisting of all Lie algebroid isomorphisms between elements in $\fldspalg$. That is, an element $\varphi \in \morphalg$ is a map $\varphi: (A,\omega) \rightarrow (A,\omega')$ which is a vector space automorphism of $A$, and moreover which preserves the brackets defined on $(A,\omega)$ and $(A,\omega')$, respectively. In \cite{Klinger:2023qna}, it is demonstrated that such maps correspond precisely with the combined set of local gauge transformations \emph{and} diffeomorphisms of the base $M$. This can be understood via the fact that $\varphi$ preserves brackets which implies that $\omega$ and $\omega'$ share the same curvature and are therefore gauge equivalent in the conventional sense. The map $\varphi$ also induces an isomorphism $g_{\varphi}^{(E)}: E \rightarrow E$ for all vector bundles associated to $A$ via a Lie algebroid representation induced by the connection reform $\omega$ \cite{Klinger:2023qna}. Given a generic element $\psi \in \Omega^{p}(A;E)$ we represent the action of $\varphi \in \morphalg$ via the Lie algebroid pullback
\beq \label{Gauge Transform}
	\psi \mapsto \varphi^* \psi, \qquad (\varphi^*\psi)\big(\mX_1, ..., \mX_p\big) = (g_{\varphi}^{(E)})^{-1}\bigg(\psi\big(\varphi(\mX_1), ... \varphi(\mX_p)\big)\bigg). 
\eeq
Although it is written in a very formal way, when pulled apart \eqref{Gauge Transform} reduces to the usual equations describing gauge transformations and diffeomorphisms of the relevant fields, as was described in detail in \cite{Klinger:2023qna,Jia:2023tki}. These together combine into a section of the Lie algebroid, and in the following we will refer to such as generalized gauge transformations.

The sets $\fldspalg$ and $\morphalg$ provide the necessary kinematical data for constructing the extended phase space of an arbitrary gauge theory. The field content of such a theory is given by the configuration data $\Phi = (\Phi^1, ..., \Phi^N)$, where each individual field $\Phi^i \in \Omega^{p_i}(A;E_i)$ is a $p_i$ form taking values in a representation $E_i$. The fields $\Phi$ have momentum $\Pi = (\Pi_1, ..., \Pi_N)$ where $\Pi_i \in \Omega^{d-1-p_i}(A;E_i^*)$ takes values in the representation dual to their conjugate field. To complete the specification of the theory we require a codimension one Cauchy surface $\Sigma \subset M$ which is embedded into $M$ via a map $\overline{\varphi}_{(1)}: \Sigma \hookrightarrow M$. More completely, we extend $\overline{\varphi}_{(1)}$ to the Lie algebroid isomorphism $\varphi_{(1)}: A_{\Sigma} \rightarrow A$, where $A_{\Sigma}$ is a Lie algebroid over a tubular neighborhood of $\Sigma$ inside $M$, $M_{\Sigma}$. 

We regard $(\Phi,\Pi)$ as defining coordinates for a symplectic manifold $X$, endowed with the canonical symplectic potential density:
\beq \label{Non-extended symplectic potential}
	\theta_{\Sigma} \equiv \varphi_{(1)}^*\bigg(\Pi_i \wedge_{A} \delta \Phi^i\bigg) \in \Omega^1(X;\Omega^{d-1}(A_\Sigma)),
\eeq 
giving rise to the symplectic potential \cite{Klinger:2023qna}:
\beq
	\Theta_{\Sigma} \equiv \int_{\Sigma} \theta_{\Sigma}. 
\eeq
To complete the specification of the theory we must also dictate dynamical data in the form of a Hamiltonian function which generates time translations. Let $\mathcal{H}: X \rightarrow \Omega^{d-1}(A)$ be a $(d-1)$-form functional of the fields which we refer to as the Hamiltonian density. Integrating,
\beq
	H_{\Sigma} \equiv \int_{\Sigma} \varphi_{(1)}^*(\mathcal{H}),
\eeq
we obtain the Hamiltonian -- regarded as a function on the phase space of the theory, $H \in C^{\infty}(X)$. Given a curve $\gamma: [0,1] \rightarrow X$ we can define the action functional in the usual way as
\begin{flalign} \label{Definition of the action}
	S &\equiv \int_{0}^{1} \bigg( \gamma^*\Theta_{\Sigma} - H_{\Sigma} \circ \gamma dt\bigg) \\
	&= \int_{\Sigma \times [0,1]} \gamma^*\varphi_{(1)}^*\bigg(\Pi_i \wedge_{A} \delta \Phi^i - \mathcal{H} dt\bigg) \equiv \int_{\Sigma \times [0,1]} \Lagr.
\end{flalign}
In \eqref{Definition of the action} we have defined the Lagrangian density $\Lagr \in \Omega^{d}(A_{\Sigma})$. 

As was introduced in \cite{Ciambelli:2021nmv,Klinger:2023qna}, in theories with gravity \eqref{Non-extended symplectic potential} must be augmented to properly account for the fact that geometrical data, such as the embedding $\varphi_{(1)}$, become part of the phase space. The extended configuration space is the principal $\morphalg$ bundle $\pbalg \rightarrow X$, with extended symplectic potential (density)\footnote{In this paper we use the notation $\hatcon{x}$ to denote the contraction by a section of an algebroid $A$. In eq. \eqref{Extended Symplectic Potential}, the contraction is with $\varpi$, regarded as a section of $A$ valued in $A^*$. See Appendix A of \cite{Klinger:2023qna} for a summary of this and related notation.}
\beq \label{Extended Symplectic Potential}
	\theta^{ext}_{\Sigma} \equiv \varphi^{*}_{(1)}\bigg(\Pi_{i} \wedge_A \delta \Phi^i - \hatcon{\MCb} \Lagr\bigg).
\eeq
Here, $\MCb \in A^* \otimes A$ is the Maurer-Cartan form associated with $\morphalg$ with the understanding that $A$ exponentiates to $\morphalg$ \cite{Klinger:2023qna}. The additional term in \eqref{Extended Symplectic Potential} arises from the non-invariance of the Lagrangian under gauge transformations (including diffeomorphisms). 
Let $a: \morphalg \times X \rightarrow X$ denote the action of $\morphalg$ on the phase space. 
We'll write this action as
\beq \label{Proj Xext to X}
a(\varphi,\tilde{\Phi},\tilde{\Pi})\equiv a_{\varphi}(\tilde{\Phi},\tilde{\Pi})=	(\Phi,\Pi).
\eeq
Alternatively, recognizing that locally $\pbalg\simeq \morphalg\times X$,  we can regard this map as a projection from $\pbalg$ with local coordinates $(\varphi,\tilde{\Phi},\tilde{\Pi})$ into $X$ with coordinates $(\Phi,\Pi)$.  


Using \eqref{Proj Xext to X} we can compute:
\beq \label{Pullback field variations}
	\delta(\Phi,\Pi) = (\delta \tilde{\Phi} + \delta_{\MCb} \tilde{\Phi}, \delta \tilde{\Pi} + \delta_{\MCb} \tilde{\Pi}),
\eeq
where $\delta_{\mX} \Phi = \hatcon{\mX} \delta_{\MCb}\Phi$ is the variation of $\Phi$ under a {generalized} gauge transformation generated by $\mX \in \Gamma(A)$. One may interpret the terms appearing in \eqref{Pullback field variations} as quantifying the free and gauge variations of the field degrees of freedom. Plugging \eqref{Pullback field variations} into \eqref{Extended Symplectic Potential} we obtain a new formulation of the extended symplectic potential:
\beq
	\theta^{ext}_{\Sigma} = \varphi_{(1)}^{*}\bigg(\tilde{\Pi}_i \wedge_A \delta \tilde{\Phi}^i + \tilde{\Pi}_i \wedge_A \delta_{\MCb} \tilde{\Phi}^{i} - \hat{i}_{\MCb} \tilde{\Lagr}\bigg).
\eeq 
Essentially, we have just pulled apart the variation of $\Phi$ into its two aforementioned pieces. The notation $\tilde{\Lagr}$ means the Lagrangian evaluated at $(\tilde{\Phi},\tilde{\Pi})$. The symplectic potential has now been divided into two pieces -- a free variational term
\beq\label{extthetaF}
	\theta^{ext}_{F} \equiv \varphi^*_{(1)}\bigg( \tilde{\Pi}_{i} \wedge_A \delta \tilde{\Phi}^i\bigg),
\eeq	
and a $\morphalg$ variational term:
\beq \label{G variational term}
	\theta^{ext}_{\morphalg} \equiv \varphi^*_{(1)}\bigg(\tilde{\Pi}_{i} \wedge_A \delta_{\MCb} \tilde{\Phi}^i - \hatcon{\MCb} \tilde{\Lagr}\bigg). 
\eeq
We note that the extended symplectic potential contains a term, eq. \eqref{extthetaF}, of the unextended phase space, written in the tilded variables, plus terms \eqref{G variational term} that correspond to the extension. This general decomposition of the extended symplectic potential is a central property of gauge theories, as we will see, and facilitates the implementation of constraints and gauge fixing. 
A significant feature of \eqref{G variational term} is that if we contract it with a vector field on phase space, we obtain the Noether current density. In the notation of \cite{Klinger:2023qna}, this is the statement that
\beq \label{Noether current}
	\algcon{-\alginjo(\mX)}\theta^{ext}_{\morphalg} = \varphi^*_{(1)}\bigg(\tilde{\Pi}_i \wedge_A \delta_{\mX} \tilde{\Phi}^i - \hatcon{\mX}\tilde{\Lagr}\bigg) \equiv \algcurrdens{\mX},
\eeq
is precisely the Noether current density. Here $-\alginjo(\mX)$ is the vertical vector field in the extended phase space that reproduces the action of $\mX\in A$ (containing locally a spacetime vector field and a Lie algebra element), with $A$ regarded as the isotropy bundle of the configuration algebroid. The Noether currents are conserved in the sense that
$\hatd \algcurrdens{\mX} = 0$ where $\hatd$ is the algebroid exterior derivative.  
The current moreover splits into two pieces:
\beq \label{curr = const + chg}
	\algcurrdens{\mX} = \constr{\mX} + \hatd \chgdens{\mX},
\eeq
where $\constr{\mX} \in H^{d-1}(A)$ and $\chgdens{\mX} \in \Omega^{d-2}(A)$. By Noether's second theorem $\constr{\mX}$ must vanish when the constraints of the theory are implemented. With that being said, $\chgdens{\mX}$ are left unconstrained and source non-zero charges supported on surfaces of codimension $2$ relative to bulk spacetime. These charges generate physical transformations and thus must not be removed in the process of constraint quantization/symplectic reduction. 

The structure of eq. \eqref{curr = const + chg} suggests that, given  
\eqref{Noether current}, we can rewrite the symplectic potential by defining Maurer-Cartan forms $\MCb_{(1)},\MCb_{(2)}$, viz
\beq \label{charge piece}
	\theta^{ext}_{\morphalg} \equiv \constr{M} \MCb_{(1)}^{M} + \hatd\bigg(\chgdens{M} \MCb_{(2)}^{M} \bigg),
\eeq
where $\constr{M}$, etc., are components relative to a local basis $\{\un{E}_{M}\}$ for $A$. Here we regard $\MCb_{(1)}$ and $\MCb_{(2)}$ as two separate copies of the Maurer-Cartan form which generate codimension one and codimension two supported transformations, respectively. In the symplectic sense, we read \eqref{charge piece} as dictating that the infinitesimal generators $\MCb_{(1)}^{M}$ and $\MCb_{(2)}^{M}$ are canonically dual to the currents $\constr{M}$ and $\chgdens{M}$. Then, the extended symplectic potential takes the final form
\begin{flalign}
	\theta_{\Sigma}^{ext} 
	&= \varphi^{*}_{(1)}\bigg(\tilde{\Pi}_{i} \wedge_A \delta \tilde{\Phi}^{i} + \hatd \bigg(\chgdens{M} \MCb_{(2)}^{M}\bigg) + \constr{M} \MCb_{(1)}^{M}\bigg).\label{pulledbackgentheta}
\end{flalign}
Or, once integrated over the hypersurface: 
\beq \label{Extended Symplectic Potential final}
	\Theta^{ext}_{\Sigma} = \int_{\Sigma} \theta^{ext}_{F} + \int_{\partial \Sigma} \chgdens{M} \MCb_{(2)}^{M} + \int_{\Sigma} \constr{M} \MCb_{(1)}^{M}. 
\eeq
At the level of the extended phase space, the symplectic form induced by  \eqref{Extended Symplectic Potential final} is non-degenerate. In the context of the gauge theory, as was mentioned below \eqref{curr = const + chg}, 
$\constr{M}$ will be interpreted as constraints; setting these to zero in some way renders the symplectic form degenerate. Gauge fixing will then reduce to a symplectic form that is non-degenerate on an appropriate quotient space. 

\subsection{Quantization of the Extended Phase Space} \label{sec: QEPS}

In this note, we take the point of view that the procedure of quantization entails the promotion of the Poisson algebra associated with a classical symplectic/Poisson geometry into a von Neumann algebra. In this section, we recall the insights of \cite{Klinger:2023tgi} to contextualize the operators which belong to the quantization of the extended phase space. In particular, we propose that the resulting von Neumann algebra be interpreted as the result of a pair of crossed products which introduce, respectively, codimension one operators that allow for the implementation of constraints and  codimension two operators implicated by the presence of a bounded subregion.  

To preface this discussion, let us first recall the correspondence between the extended phase space and the crossed product in general terms. Let $(X,\Omega)$ denote a (pre)-symplectic manifold which admits a $G$-action $a: G \times X \rightarrow X$. 
We work in the context that the $G$-action is presymplectic, meaning that it preserves the symplectic form
\beq \label{presym action}
	a_g^* \Omega = \Omega, \;\; \forall g \in G.
\eeq
An immediate corollary of \eqref{presym action} is that $a$ induces an automorphism of the Poisson algebra, $\mathcal{M}^{pq}_X$, associated with $(X,\Omega)$.\footnote{The superscript ``pq" stands for pre-quantum to emphasize that the Poisson algebra has not yet been quantized.} Nonetheless, it is not manifest that the generators of the $G$-action can themselves be represented by elements of the Poisson algebra, rather it is merely true that   $G$ acts on the Poisson algebra as an automorphism. For this reason, it is not immediately clear that the generators of  $G$ can be quantized using standard methods. Conversely, if the algebra of generators of the $G$ action \emph{are} faithfully encoded in the Poisson algebra it is termed \emph{equivariant}. In other words, there exists an algebra homomorphism
\beq
	\Phi: \mathfrak{g} \rightarrow \mathcal{M}^{pq}_{X},
\eeq
called a \emph{moment map}. The quantization of such a moment map is well understood. Thus, lifting a presymplectic action to an equivariant one is a useful step in completing the quantization of a theory admitting a group action.

In \cite{Klinger:2023tgi} we demonstrated that, starting from a (pre)-symplectic manifold $(X,\Omega)$ with presymplectic $G$-action $a$, there exists a principal $G$-bundle, $X_{ext} \rightarrow X$, for which the presymplectic $G$-action can be extended to an equivariant $G$-action. (In the last section, the role of $X_{ext}$ was played by $\pbalg$). In other words, the Poisson algebra $\mathcal{M}^{pq}_{X_{ext}}$ always admits a moment map $\chg{}: \mathfrak{g} \rightarrow \mathcal{M}^{pq}_{X_{ext}}$. The usefulness of $X_{ext}$ is that it provides a prequantum phase space complete with the generators of the group $G$ which can, at least in principle, be quantized. 

The crux of the connection between the extended phase space and the crossed product is that the above procedure can be understood as a (pre)-symplectic analog of the following scenario in the context of von Neumann algebras. Let $\mathcal{M}$ be a von Neumann algebra acted upon by the $G$-automorphism $\alpha: G \times \mathcal{M} \rightarrow \mathcal{M}$. Moreover, let $\pi: \mathcal{M} \rightarrow B({\cal H})$ denote a faithful representation of the algebra on a Hilbert space ${\cal H}$ which also carries a covariant representation of the triple $(\mathcal{M},G,\alpha)$. Then, there exists a unitary representation $U: G \rightarrow U({\cal H})$ such that
\beq
	\pi \circ \alpha_g(x) = U(g) \pi(x) U(g)^{\dagger}, \qquad x \in \mathcal{M}, g \in G.
\eeq
The $G$-automorphism $\alpha$ is termed \emph{inner} if and only if $\text{im}(G) \subset \text{im}(\pi)$; i.e., if and only if the group elements can be understood as operators inside of the intrsinsic algebra $\mathcal{M}$. Otherwise, the automorphism is termed \emph{outer}. The situation of having an outer automorphism is akin to having a presymplectic but not equivariant $G$-action at the prequantum level. As the extended phase space provides a mechanism for promoting presymplectic actions to equivariant actions, the crossed product provides a mechanism for promoting outer automorphisms to inner ones. Explicitly, the crossed product can be understood as the von Neumann algebra generated by combining the images of the maps $\pi$ and $U$:
\beq
	\mathcal{M} \rtimes_{\alpha} G \equiv \{\text{im}(\pi) \otimes \text{im}(U)\}''. 
\eeq
The basic argument of \cite{Klinger:2023tgi} is that a suitable quantization of the Poisson algebra of the extended phase space results in a von Neumann algebra that is isomorphic to the crossed product of $\mathcal{M}^{pq}_X$ and the group $G$ with respect to an appropriate automorphism induced by the $G$-action $a$. 

Having set the stage, we can now describe the role of the extended phase space/crossed product construction as it pertains to the quantization of gauge theories on subregions. The starting point, as we have discussed, is the symplectic geometry induced by the extended phase space: $(\pbalg, \sym_{ext})$. This geometry is acted upon equivariantly by the groupoid of morphisms $\morphalg$, which corresponds to the set of all local gauge transformations \emph{and} diffeomorphisms. In particular, this implies that there exists an equivariant moment map $\chg{}$ and thus, at least in principle, this symplectic geometry can be  quantized. This quantization promotes each of the Hamiltonian charge functions to operators which we signify by the addition of a hat: $\chg{} \mapsto \hat{\chg{}}$. We denote by $\mathcal{M}_{\pbalg}$ the von Neumann algebra obtained by quantizing the Poisson algebra associated with $\pbalg$.  

Recall that $\pbalg$ is a principal $\morphalg$-bundle over $X$, the symplectic manifold associated with the field data alone. Let us dwell for a moment on $X$. As is well known \cite{de1998lie}, $X$ itself has the structure of a principal $\morphalg$-bundle over the quotient space $X/\morphalg$. In other words, we may regard the space of all field configurations as the set of all gauge \emph{inequivalent} field configurations along with their $\morphalg$-orbits. To be explicit, the copy of $\morphalg$ which appears in $X$ is generated by codimension one currents $\constr{\mX}$ as  in \eqref{curr = const + chg}. In other words, the extended phase space has the following schematic structure:
\beq \label{Double Extension}
	\pbalg \rightarrow X \rightarrow X/\morphalg_{(1)}.
\eeq

We interpret \eqref{Double Extension} as identifying the phase space $\pbalg$ as the result of \emph{two} extensions -- one with respect to the group of constraints $\morphalg_{(1)}$ and one with respect to the group of physical charges $\morphalg_{(2)}$. Reiterating an earlier discussion, each charge splits into two pieces: 
\beq \label{Charge split}
	\chg{\mX} \equiv \int_{\Sigma} \constr{\mX} + \int_{\partial \Sigma} \chgdens{\mX} \equiv \chgcon{\mX} + \chgchg{\mX}. 
\eeq
Hereafter we will refer to the former as \emph{constraint charges}, and to the latter as the \emph{physical charges}. We regard $\chgcon{}: A \rightarrow \Omega^0(\pbalg)$ and $\chgchg{}: A \rightarrow \Omega^0(\pbalg)$ as generating two separate representations of the gauge algebra $A$ in the Poisson algebra of $\pbalg$. Let ${\cal H}_{aux}$ be a Hilbert space for which the algebra $\mathcal{M}_{\pbalg}$ possesses a faithful representation $\pi: \mathcal{M}_{\pbalg} \rightarrow B({\cal H}_{aux})$. Then, at the level of ${\cal H}_{aux}$, \eqref{Charge split} implies that we have a pair of unitary representations:
\beq
	U^{(1)}\bigg(\text{exp}(\mX)\bigg) = \text{exp}\bigg(\hat{\mathbb{H}}^{(1)}_{\mX}\bigg), \qquad U^{(2)}\bigg(\text{exp}(\mX)\bigg) = \text{exp}\bigg(\hat{\mathbb{H}}^{(2)}_{\mX}\bigg),
\eeq 
which give rise to two separate von Neumann algebra automorphisms
\begin{flalign}
	&\pi \circ \alpha_{\text{exp}(\mX)}^{(1)}(x) = U^{(1)}\bigg(\text{exp}(\mX)\bigg) \pi(x) U^{(1)}\bigg(\text{exp}(\mX)\bigg)^{\dagger}, \\
	&\pi \circ \alpha_{\text{exp}(\mX)}^{(2)}(x) = U^{(2)}\bigg(\text{exp}(\mX)\bigg) \pi(x) U^{(2)}\bigg(\text{exp}(\mX)\bigg)^{\dagger}.
\end{flalign}
The von Neumann algebra resulting from the quantization of the extended phase space is therefore of the form
\beq \label{Double Cross}
	\mathcal{M}_{\pbalg} \simeq \mathcal{M}_{X} \rtimes_{\alpha^{(2)}} \morphalg_{(2)} \rtimes_{\alpha^{(1)}} \morphalg_{(1)}. 
\eeq
Eqn. \eqref{Double Cross} is the von Neumann algebra analog of \eqref{Double Extension}. 

The conclusion of the preceding paragraph could have qualitatively been obtained by observing the symplectic potential \eqref{Extended Symplectic Potential final}, which designates the classical variables that should be quantized to operators. From that perspective, the complete quantum theory should have operators associated with three sets of conjugate pairs: $(\tilde{\Phi},\tilde{\Pi})$ corresponding to (dressed) field configuration data, $(\MCb_{(1)},\constr{})$ corresponding to generators and charges associated with pure gauge transformations, and $(\MCb_{(2)},\chgdens{})$ corresponding to generators and charges associated with physical transformations supported on corners. One may therefore regard the operator algebra realized by quantizing the extended phase space as
\beq \label{Skeleton of Extended Algebra}
	\mathcal{M}_{\pbalg} \sim \{ \hat{\tilde{\Phi}},\hat{\tilde{\Pi}}, \hat{\MCb}_{(1)},\hat{\constr{}}, \hat{\MCb}_{(2)},\hat{\chgdens{}} \}''.
\eeq

To summarize this section, we have now recognized how the extended phase space prepares a quantum theory in which all of the relevant operators are present including operators which correspond with constraints. In the remainder of this note we will discuss the naturally related problem of implementing these aforementioned constraints in a manner which is quantum mechanically consistent. 

\section{Constraints Four Ways} \label{sec: con-quant}

In this section we will demonstrate how the extended phase space figures naturally in the implementation of constraints by considering four distinct methods: (i) Refined Algebraic Quantization, (ii) BRST Quantization, (iii) Path Integral Quantization, and (iv) the commutation theorem. In this respect, we will also seek to accomplish the goal of unifying these distinct approaches to quantizing systems with constraints under the umbrella of a single geometric framework. As we will demonstrate, all four approaches to constraint quantization are linked in the sense that they construct projection maps originating from the crossed product algebra. 

In what follows we will concentrate on generic algebras of the form
\beq \label{Starting Point}
	\cpalg \equiv \mathcal{M} \rtimes_{\alpha} G.
\eeq
We may regard $\cpalg$ as arising from the  quantization of an extended phase space $X_{ext} \rightarrow X$, where $(X,\Omega)$ is a symplectic manifold acted upon at least presymplectically by the group $G$.\footnote{Since our analysis is primarily aimed at understanding gauge theories, it is more rigorously correct to regard the ``group" $G$ as having the structure of a Lie groupoid -- as was implicit in the discussion in Section \ref{sec: preliminaries}. To avoid overloading the discussion with too much technical detail, we will often refer to $G$ merely as a group. For a discussion of crossed products with locally compact topological groupoids see \cite{paterson2012groupoids,landsman1999lie}.} In this regard it will also be beneficial to regard $\mathcal{M}$ as the  algebra obtained by quantizing the non-extended space $X$. Our goal will be to describe how the machinery of the crossed product allows for the seamless implementation of constraints associated with the group $G$. Let us emphasize, in light of Section \ref{sec: QEPS}, that when we have a gauge theory in mind this does \emph{not} entail constraining the physical charges in any way. One may always think of the gauge theory algebra as being of the form
\beq
	\mathcal{M}_{\pbalg} \simeq \bigg(\mathcal{M}_{X} \rtimes_{\alpha^{(2)}} \morphalg_{(2)} \bigg) \rtimes_{\alpha^{(1)}} \morphalg_{(1)},
\eeq
where the constraint quantization procedures described below are performed for $\morphalg_{(1)}$ while treating $\mathcal{M}_{X} \rtimes_{\alpha^{(2)}} \morphalg_{(2)}$ as its own algebra. This observation plays out in the examples discussed in Section \ref{sec: Zamples}. 

Before moving into the analysis of this section, it will be prudent to first review different kinds of maps accessible in the operator algebraic context which share properties we associate with conditional expectations. For the purpose of this discussion, let $M$ be a von Neumann algebra and $N \subset M$ a von Neumann subalgebra. We will also use the notation $\hat{M}$ to denote the set of operators affiliated with a von Neumann algebra $M$. Given a faithful representation $\pi: M \rightarrow B(H)$, a closed operator $\mathcal{O}$ on $H$ is called \emph{affiliated} to $M$ if $\mathcal{O}$ commutes with all unitary elements in $M'$. If $\mathcal{O}$ is bounded and affiliated to $M$ then it is in $M$ by von Neumann's theorem. 

An \emph{operator-valued weight} \cite{haagerup1979operator,haagerup1979operator2} is a map $T: M_+ \rightarrow \hat{N}$ which is linear and satisfies the bi-module property
\beq \label{Bi-module property}
	T(n_1^* m n_2) = n_1^* T(m) n_2, \; \forall m \in M, n_1,n_2 \in N \subset M.
\eeq
It is worth noting that, in the case that $N = \mathbb{C} \mathbb{1} \subset M$, the definition of an operator-valued weight reduces to that of an ordinary weight. As is the case for ordinary weights, the map $T$ defining an operator-valued weight can be extended uniquely to a map $T: M \rightarrow \hat{N}$ by decomposing elements in $M$ into the difference of its positive and negative parts. If an operator-valued weight $T$ is unital, i.e. $T(\mathbb{1}_M) = \mathbb{1}_N$, it is called a \emph{conditional expectation}. In the case that $N = \mathbb{C}\mathbb{1}$, a conditional expectation defines what is typically referred to as a state. Given a state $\varphi$ on $M$, we say that the conditional expectation $E: M \rightarrow N$ preserves the state if
\beq
	\varphi = \varphi\rvert_N \circ E. 
\eeq
This is the non-commutative analog of factorizing a probability density as a product of a marginal and a conditional density. A theorem by Takesaki \cite{takesaki1972conditional} states that a state preserving conditional expectation will only exist if $N$ is invariant under the modular automorphism generated by $\varphi$. 

Alternatively to operator-valued weights one has \emph{generalized conditional expectations} \cite{accardi1982conditional}. To introduce the generalized conditional expectation, let us first recall the Petz dual of a quantum channel. Given a state $\varphi$ on $M$ the KMS inner product is defined (up to a possible quotient by its kernel) by
\beq	 \label{KMS inner}
	g^{KMS}_{M,\varphi}(m_1,m_2) \equiv \varphi\bigg(m_1^* \sigma^{\varphi}_{-i/2}(m_2)\bigg), \; m_1, m_2 \in M.
\eeq 
In \eqref{KMS inner} $\sigma^{\varphi}: \mathbb{R} \rightarrow \text{Aut}(M)$ is the modular automorphism induced by $\varphi$. Next, let $\alpha: M \rightarrow N$ be a normal, completely positive, unital map between two von Neumann algebras (for now $N$ need not be a subalgebra of $M$). Such a map is called a \emph{quantum channel}. Let $\varphi_0$ be a normal state on $N$ such that $\varphi \equiv \varphi_0 \circ \alpha$ is a normal and faithful state on $M$. Then, we define the \emph{Petz dual} \cite{ohya2004quantum} of the channel $\alpha$ with respect to the state $\varphi_0$ as the map $\alpha^{\dagger}_{\varphi_0}: N \rightarrow M$ which is formally adjoint to $\alpha$ with respect  to the intertwining inner products $g^{KMS}_{N,\varphi_0}$ and $g^{KMS}_{M,\varphi_0 \circ \alpha}$. That is,
\beq \label{Petz Dual}
	g^{KMS}_{M,\varphi_0 \circ \alpha}\bigg(m,\alpha^{\dagger}_{\varphi_0}(n)\bigg) = g^{KMS}_{N,\varphi_0}\bigg(\alpha(m),n\bigg), \; \forall m \in M, n \in N. 
\eeq
It can be shown that $\alpha^{\dagger}_{\varphi_0}$ is also a quantum channel. 

In the case that $N \subset M$ is a von Neumann subalgebra, the inclusion $i: N \hookrightarrow M$ is a normal, completely positive, unital map and given any state $\varphi$ on $M$ the state $\varphi \circ i$ is normal and faithful on $N$. Thus, there exists a canonical quantum channel $E^{N}_{\varphi}: M \rightarrow N$ obtained from the Petz dual of $i$ with respect to the state $\varphi$. The map $E^N_{\varphi}$ is called the \emph{generalized conditional expectation} induced by the pair $(N,\varphi)$. The map $E^N_{\varphi}$ is always unital and preserves the state $\varphi$ in the sense that
\beq
	\varphi = \varphi\rvert_{N} \circ E^N_{\varphi}. 
\eeq
However, the generalized conditional expectation does not automatically possess the bi-module property of an operator-valued weight. This fact is expressed in the following theorem \cite{accardi1982conditional}: Given $n \in N$ the following are all equivalent:
\begin{enumerate}
	\item $E^N_{\varphi}(nm) = n E^N_{\varphi}(m), \; \forall m \in M$,
	\item $\sigma^{\varphi}_t(n) \in N$, 
	\item $\sigma^{\varphi\rvert_N}_t(n) = \sigma^{\varphi}_t(n), \; \forall t$. 
\end{enumerate}
In other words, the generalized conditional expectation obtains the bi-module property provided the modular automorphism of $\varphi$, restricted to $N$, is the same as the modular automorphism of $\varphi\rvert_N$. In the event that $E^N_{\varphi}$ possesses the bi-module property for all $n \in N$, it can be shown that the generalized conditional expectation is equivalent to the unique $\varphi$-preserving conditional expectation from $M$ to $N$.

In summary, the strict definition of a conditional expectation is a linear, unital map $E: M \rightarrow N$ satisfying the bi-module property \eqref{Bi-module property}. In general the existence of a conditional expectation between an algebra and a given subalgebra is not guaranteed. However, one can always realize conditional expectation `like' maps by relaxing either the unital condition -- in which case one obtains an operator-valued weight -- or the bi-module condition -- in which one obtains a generalized conditional expectation. In the following we will find need to use these relaxations of the conditional expectation to construct projection maps which implement constraints. 

\subsection{Refined Algebraic Quantization} \label{sec: RAQ}

The starting point of Refined Algebraic Quantization (RAQ) is a Hilbert space $\mathcal{H}_{aux}$ upon which the group of constraints $G$ are realized as unitary operators. One way of thinking about $\mathcal{H}_{aux}$ is that it corresponds to the quantization of the \emph{unconstrained} system associated with the \emph{constrained} system we are interested in studying. As was alluded to in Section \ref{sec: preliminaries}, this is precisely the role of the extended phase space. 

More explicitly, starting from the crossed product \eqref{Starting Point} there is a canonical Hilbert space which can play the role of $\mathcal{H}_{aux}$.\footnote{More generally, the role of the auxiliary Hilbert space can be played by any covariant representation of $\cpalg$. Given a Hilbert space $\mathcal{H}_{cov}$, a \emph{covariant representation} is a pair $(\pi,\lambda)$ where $\pi: \mathcal{M} \rightarrow B(\mathcal{H}_{cov})$ is a faithful representation of $\mathcal{M}$, $\lambda: G \rightarrow U(\mathcal{H}_{cov})$ is a unitary representation of $G$ and these representations are compatible with $\alpha$ in the sense that
\beq
	\pi \circ \alpha_g(x) = \lambda(g) \pi(x) \lambda(g)^*.
\eeq} Let $\mathcal{H}$ be any Hilbert space for which there exists a faithful representation $\pi: \mathcal{M} \rightarrow B(\mathcal{H})$. Then, as is established in \cite{takesaki1973crossed}, we can realize the crossed product algebra as operators acting on an extended Hilbert space equivalent to the set of square integrable functions on the group $G$ with values in the Hilbert space $\mathcal{H}$, $\mathcal{H}_{aux} = L^2(\mathcal{H};G,\mu) \simeq L^2(G) \otimes \mathcal{H}$. Given two such elements, $\xi, \xi' \in \mathcal{H}_{aux}$, we define the inner product
\beq \label{ext0 inner}
	K_{aux}(\xi,\eta) \equiv \int_G \; \mu(g) \; h\big(\xi(g), \xi'(g)\big),
\eeq
where here $\mu(g)$ is the left-invariant Haar measure on $G$, and $h: \mathcal{H} \times \mathcal{H} \rightarrow \mathbb{C}$ is an inner product on $\mathcal{H}$.\footnote{This analysis can be generalized rather straightforwardly to the case of a Lie groupoid by passing from a discussion of left invariant Haar measures for Lie groups to left invariant Haar systems for locally compact topological groupoids.}
 We may subsequently define the explicit representations:
\beq \label{ext0 rep}
	\bigg( \pi_{\alpha}(x) \big(\xi\big)\bigg)(g) \equiv \pi \circ \alpha_{g^{-1}}(x) \big( \xi(g) \big), \qquad 
	\bigg( U(g') \big(\xi\big) \bigg)(g) \equiv \xi(g'^{-1} g),
	 \qquad x \in \mathcal{M},\; g,g' \in G.
\eeq
The representation $U: G \rightarrow B(\mathcal{H}_{aux})$ is unitary by construction due to the left invariance of \eqref{ext0 inner}. Occasionally it may be necessary to notate the representation of an arbitrary element $x \in \cpalg$ without specifying if it is in $\mathcal{M}$ or $G$. In these cases we write $\rho(x)$ and let the explicit representation be dictated by the nature of the element $x$.

From the quantum mechanical perspective implementing the constraints of the theory corresponds to identifying a Hilbert space of elements which are left invariant by the constraint operators -- the so-called physical Hilbert space $\mathcal{H}_{phy
s}$. RAQ provides a direct construction of this space. The impetus for RAQ was an observation that Dirac's approach to constraint quantization \cite{dirac1964lectures} is ill-fated because physical states typically don't lie inside of the Hilbert space one starts with under the naive quantization of a constrained system \cite{Ashtekar:1995zh,Giulini:1998kf,Giulini:1998rk,Marolf:2000iq,Marolf:1996gb}. Rather, it is necessary to consider an enlarged Hilbert space in which the physical states are ``filled back in". 

To accomplish this aim, we let $\mathcal{H}_{aux}^*$ denote the algebraic dual of $\mathcal{H}_{aux}$ endowed with the topology of pointwise convergence.\footnote{That is, $\{f_n\} \subset \mathcal{H}_{aux}^*$ converges to $f \in \mathcal{H}_{aux}^*$ if and only if the sequences $\{f_n(\un{\phi})\} \subset \mathbb{C}$ converge to $f(\un{\phi})$ for each $\un{\phi} \in \mathcal{H}_{aux}$.} The representation \eqref{ext0 rep} descends to a representation on $\mathcal{H}_{aux}^*$ via the dual action $U_*: G \rightarrow B(\mathcal{H}_{aux}^*)$,\footnote{Passing to the dual space circumvents possible problems with existence of solutions to $U(g)\un\psi=\un\psi$ \cite{Ashtekar:1995zh,Marolf:2000iq}. We will not dwell on the more technical details in the main text, however the reader can find a more complete discussion of rigging in Appendix \ref{app: rigging}.}
\beq \label{Dual Action}
\bigg(U_*(g)\big(f\big)\bigg)(\xi) = f\bigg(U(g^{-1})\big(\xi\big)\bigg),\qquad f\in \mathcal{H}_{aux}^*, g\in G,\xi\in \mathcal{H}_{aux}.
\eeq
More generally, the dual action of $\cpalg$ on $\mathcal{H}_{aux}^*$ is given by
\beq
	\bigg(\rho_*(x)\big(f\big)\bigg)(\xi) \equiv f\bigg(\rho(x)^{\dagger}\big(\xi\big)\bigg). 
\eeq

We can now construct a set of states in $\mathcal{H}_{aux}^*$ which are invariant under the action \eqref{Dual Action}. Such states are realized by defining a so-called \emph{rigging map} \cite{Landsman:1993xe} $\eta: \mathcal{H}_{aux} \rightarrow \mathcal{H}_{aux}^*$ satisfying the following properties:
\begin{eqnarray}
\label{Invariance}  &\bigg(U_*(g)\big(\eta(\xi_1)\big)\bigg)(\xi_2) = \eta(\xi_1)\bigg(U(g)\big(\xi_2\big)\bigg) = \eta(\xi_1)\big(\xi_2\big), \qquad \forall \xi_1,\xi_2 \in \mathcal{H}_{aux}, g \in G, &\\
\label{Hermiticity} &\eta(\xi_1)\big(\xi_2\big) = \overline{\eta(\xi_2)\big(\xi_1\big)}, \qquad \forall \xi_1, \xi_2 \in \mathcal{H}_{aux},& \\
\label{Positivity} &\eta(\xi)\big(\xi\big) \geq 0, \qquad \forall \xi\in \mathcal{H}_{aux}.&
\end{eqnarray}
Condition \eqref{Invariance} implies that the image of $\eta$ are invariant under the dual action of $G$. Conditions \eqref{Hermiticity} and \eqref{Positivity} together imply that
\beq \label{physical inner}
	h_{phys}\bigg(\eta(\xi_1),\eta(\xi_2)\bigg) \equiv \eta(\xi_1)\big(\xi_2\big),
\eeq
defines an inner product on $\eta(\mathcal{H}_{aux}) \subset \mathcal{H}_{aux}^*$; the first implying that \eqref{physical inner} is sesquilinear and the second implying that it is positive definite. Thus, conditions \eqref{Invariance}-\eqref{Positivity} imply that we can define a physical Hilbert space by taking the closure of $\eta(\mathcal{H}_{aux}) \subset \mathcal{H}_{aux}^*$ under the topology induced by the inner product \eqref{physical inner}: $\mathcal{H}_{phys} \equiv \overline{\eta(\mathcal{H}_{aux})}$. 

In \cite{Giulini:1998kf,Giulini:1998rk} an explicit form for a rigging map satisfying the aforementioned properties was given in terms of a group-averaging procedure. Moreover, it was shown that whenever such a procedure is well-defined, the resulting map is the unique rigging map up to a multiplicative constant. For simplicity, we assume that $G$ is a unimodular group meaning it possesses a unique bi-invariant Haar measure $\mu$.\footnote{If this is not the case, i.e., if $G$ possesses distinct left- and right-invariant Haar measures one can utilize the procedure of unimodularization to generalize the following discussion. See \cite{Giulini:1998kf}.} Then, we define the rigging map
\beq \label{Group averaging}
	\eta(\xi_1)\big(\xi_2\big) = K_{aux}\bigg(\xi_2, \int_G \mu(g) \; U(g)\big(\xi_1\big)\bigg).
\eeq
It is straightforward to show that \eqref{Group averaging} satisfies conditions \eqref{Invariance}-\eqref{Positivity}. Most crucially, the invariance of $\eta(\xi)$ follows immediately from the invariance of the Haar measure:
\begin{flalign}
	\bigg(U_*(g)\big(\eta(\xi_1)\big)\bigg)(\xi_2) &= \eta(\xi_1)\bigg(U(g^{-1})\big(\xi_2\big)\bigg) \nonumber \\
	&= K_{aux}\bigg(U(g^{-1})\big(\xi_2\big), \int_G \mu(g') \; U(g')\big(\xi_1\big)  \bigg) \nonumber \\
	&= K_{aux}\bigg( \xi_2, \int_G \mu(g') \; U(gg') \big(\xi_1 \big)  \bigg) \nonumber \\
	&= K_{aux}\bigg( \xi_2, \int_G \mu(k) \; U(k)\big(\xi_1\big)  \bigg) = \eta(\xi_1)\big(\xi_2\big). 
\end{flalign}

In the preceding discussion, we have stressed a Hilbert space oriented approach to implementing constraints, as is typical in the literature. However, as we will now describe, there is a very natural alternative approach to the problem of constructing the physical space of states which leads instead with the algebra. To begin, let us define by
\beq \label{Invariant Algebra}
	\mathcal{M}_{inv} \equiv \{ x \in \cpalg \; | \; \alpha_g(x) = x \; \forall g \in G\}
\eeq
the set of $G$-invariant operators in $\cpalg$. Moreover, let $\omega: \mathcal{M}_{inv} \rightarrow \mathbb{C}$ denote a weight on $\mathcal{M}_{inv}$. Given this data, we can form the GNS representation of $\mathcal{M}_{inv}$, $\mathcal{H}_{inv}$ with inner product $g_{\omega}$.\footnote{For a formal primer on the GNS construction, see Appendix A of \cite{Klinger:2023tgi}.} The GNS construction is characterized by the maps $\eta_{\omega}: \mathcal{M}_{inv} \rightarrow \mathcal{H}_{inv}$ and $\pi_{\omega}: \mathcal{M}_{inv} \rightarrow B(\mathcal{H}_{inv})$ along with a vector ${\psi}_{\omega} \in \mathcal{H}_{inv}$ such that
\beq \label{Characterization of GNS}
	g_{\omega}(\eta_{\omega}(x),\eta_{\omega}(y)) = \omega(x^*y), \qquad \eta_{\omega}(x) = \pi_{\omega}(x){\psi}_{\omega}. 
\eeq	
Notice that a corollary of \eqref{Characterization of GNS} is that
\beq \label{GNS State}
	\omega(x) = g_{\omega}\bigg({\psi}_{\omega}, \pi_{\omega}(x) {\psi}_{\omega}\bigg)
\eeq
which is the standard result that $\omega(x)$ computes the expectation value of $x$ in ${\psi}_{\omega}$.

The group $G$ is realized on $\mathcal{H}_{inv}$ in terms of a unitary representation $U_{\omega}: G \rightarrow U(\mathcal{H}_{inv})$ which is compatible with the automorphism $\alpha$ in the sense that $\pi_{\omega} \circ \alpha_g(x) = U_{\omega}(g) \pi_{\omega}(x) U_{\omega}(g)^{-1}$. Thus, by \eqref{Invariant Algebra}
\beq \label{GNS Commutation}
	\pi_{\omega}(x) = \pi_{\omega} \circ \alpha_g(x) = U_{\omega}(g) \pi_{\omega}(x) U_{\omega}(g)^{-1}, \text{ or } [U_{\omega}(g),\pi_{\omega}(x)] = 0.
\eeq
At the same time, using \eqref{GNS State} along with \eqref{Invariant Algebra} we can write
\beq
	\omega(x) = g_{\omega}\bigg({\psi}_{\omega}, U_{\omega}(g) \pi_{\omega}(x) U_{\omega}(g)^{-1} {\psi}_{\omega} \bigg) = g_{\omega}\bigg(U_{\omega}(g)^{-1}{\psi}_{\omega}, \pi_{\omega}(x) U_{\omega}(g)^{-1}{\psi}_{\omega}\bigg), \; \forall x \in \mathcal{M}_{inv}, 
\eeq  
which implies that 
\beq \label{Invariance of GNS State}
	U_{\omega}(g){\psi}_{\omega} = {\psi}_{\omega}, \; \forall g \in G.
\eeq 
Combining \eqref{GNS Commutation} and \eqref{Invariance of GNS State}, we can easily show that each element in the GNS Hilbert space is invariant under the action of the group:
\beq
	U_{\omega}(g)(\eta_{\omega}(x)) = U_{\omega}(g) (\pi_{\omega}(x){\psi}_{\omega}) = \pi_{\omega}(x) U_{\omega}(g){\psi}_{\omega} = \pi_{\omega}({\psi}_{\omega}) = \eta_{\omega}(x). 
\eeq
Thus, we conclude that $\mathcal{H}_{inv}$ is an alternative construction of a physical Hilbert space. 

The connection between the standard approach to RAQ and the algebraic approach based on the GNS construction arises when we consider the remaining loose end, which is an identification of the algebra \eqref{Invariant Algebra}. To do so we define the map
\beq \label{Group averaging in algebra}
	\Pi_d: \cpalg \rightarrow \mathcal{M}_{inv}, \qquad x \mapsto \int_G \mu(g) \; \alpha_g(x).
\eeq
Eq. \eqref{Group averaging in algebra} is nothing but the group averaging applied directly to the algebra. It is easy enough to check that
\beq
	\alpha_g \circ \Pi_d(x) = \int_G \mu(g') \; \alpha_{gg'}(x) = \int_G \mu(k) \; \alpha_k(x) = \Pi_d(x). 
\eeq
Thus, we may take $\mathcal{M}_{inv} \equiv \Pi_d(\cpalg)$. More to the point, if we extend $\omega$ to a weight on the full algebra $\cpalg$ then we can construct a GNS representation for the full algebra which we denote by $\mathcal{H}_{cp}$ along with its corresponding set of data. In the context of the GNS Hilbert space $\mathcal{H}_{cp}$ we can construct the map
\beq \label{Rigging Map from Algebraic Perspective}
	\eta = \eta_{\omega} \circ \Pi_{d}: \cpalg \rightarrow \mathcal{H}_{cp}, \qquad x \mapsto \int_G \mu(g) \; U_{\omega}(g) \big(\eta_{\omega}(x)\big),
\eeq
which is precisely the group averaging map in $\mathcal{H}_{cp}$. 

There are some subtleties that we should address in order to ensure the validity of \eqref{Rigging Map from Algebraic Perspective}. We have tacitly assumed that $\omega$ is invariant under the action of $G$:
\beq \label{State invariance}
	\omega \circ \alpha_g = \omega, \; \forall g \in G.
\eeq	
This was true trivially when restricted to the subalgebra $\mathcal{M}_{inv}$, but it is not immediately clear that \eqref{State invariance} holds when applied to the full algebra. One way of interpreting this state of affairs is that we have reduced the problem of finding a complete Hilbert space of group invariant elements down to the problem of finding just one such element. Once this initial element, ${\psi}_{\omega}$, is identified all other such states can be constructed via \eqref{Rigging Map from Algebraic Perspective} which we interpret as the vector created by acting on ${\psi}_{\omega}$ with $G$-invariant operators. 

It is at this point that we arrive at a very important insight --- if $\cpalg$ possesses a tracial weight $\tau: \cpalg \rightarrow \mathbb{C}$, then it can always play the role of our desired $G$-invariant element. Indeed, $\tau \circ \alpha_g = \tau$ follows immediately from the cyclicity property of the trace, provided the automorphism $\alpha$ is inner unitarily implemented which the crossed product also guarantees. As was discussed in great detail in \cite{Klinger:2023tgi}, $\cpalg$ will possess such a trace if and only if it is semi-finite. The semi-finite nature of a von Neumann algebra is intimately related with other desirable properties such as the existence of density operators and entropies.  In the aforementioned paper and in other related work \cite{Witten:2021unn, Chandrasekaran:2022eqq, Chandrasekaran:2022cip,Jensen:2023yxy}, it has been suggested that a physical theory which possesses diffeomorphism covariance should always been semi-finite. This can be understood as an immediate consequence of the relationship between the extended phase space and the crossed product, in addition to the role of corner supported diffeomorphisms in specifying the modular automorphism group of a subregion algebra \cite{faulkner2016shape, faulkner2016modular}. Thus, once again, we see that the crossed product saves the day: so long as $\cpalg$ is a semi-finite von Neumann algebra, we can construct the GNS Hilbert space associated with the tracial weight and then perform the group averaging procedure implicated by \eqref{Rigging Map from Algebraic Perspective} in order to construct a complete set of physical states. 

One might be inclined to think of the tracial weight $\tau$ and its associated Hilbert space representation, ${\psi}_{\tau}$, as playing the role of the $G$-invariant vacuum. Then, the content of \eqref{Rigging Map from Algebraic Perspective} is simply that all of the physical states of the theory can be obtained by perturbing the vacuum by gauge invariant operators. This is closely related to the conclusions reached in \cite{Chakraborty:2023yed,Chakraborty:2023los} in which the authors construct a Hilbert space of states in de Sitter quantum gravity. This is unsurprising, as \cite{Chakraborty:2023yed,Chakraborty:2023los} is a modification on a classic result by Higuchi \cite{Higuchi:1991tk, Higuchi:1991tm} which produces a space of de Sitter invariant states in the non-gravitational limit precisely by group averaging over the symmetries of the background. The more recent papers refine this approach by making use of the Fadeev-Popov method for implementing constraints in the path integral sense. We will revisit the map \eqref{Group averaging in algebra} again in Section \ref{sec: Path} where it appears naturally in precisely this context. 

As we have alluded to, there is one further subtlety which warrants attention as it concerns the use of the map \eqref{Group averaging in algebra}. In the case that the group $G$ is compact, \eqref{Group averaging in algebra} is a conditional expectation as defined in Section \ref{sec: con-quant}. In general, however, the integral over the group will diverge when applied to the identity element, and thus the map $\Pi_d$ cannot be unital.\footnote{In fact, this issue is closely related to whether the tracial weight utilized above, $\tau$, is moreover a state in which case the crossed product algebra is of type II$_1$ as opposed to type II$_{\infty}$.} In this case $\Pi_d$ is only an operator-valued weight, rather than a conditional expectation. This is an issue that has plagued the RAQ approach to constraint quantization since its inception, since an operator-valued weight cannot be used to construct normalizable states. In particular, this seems to put a damper on the usefulness of $\Pi_d$ as a tool for implementing constraints in gravitational theories wherein the group of symmetries is not compact. In this regard, however, passing from the Hilbert space oriented approach to the operator algebraic approach seems serendipitous. Although an explicit form in terms of group averaging is not valid in this case, we can in principle upgrade the map $\Pi_d$ by appealing to Accardi and Cecchini's construction of the generalized conditional expectation \cite{accardi1982conditional}. This map, while lacking the bi-module property of an operator-valued weight, is manifestly unital and so may be used to construct normalizable states. Then, although the form of the maps \eqref{Group averaging in algebra} and \eqref{Rigging Map from Algebraic Perspective} may be different, the ideas of the preceding discussion  can be enhanced to make sense even in this case. In future work we plan to address the problem of constructing such a generalized conditional expectation in detail.

\subsection{BRST Quantization} \label{sec: BRST}

The next approach to implementing constraints that we would like to consider is the BRST quantization scheme. 
As was the case for the Refined Algebraic Quantization scheme, BRST quantization begins with an enlarged Hilbert space which includes states that will ultimately be regarded as unphysical. In particular, one introduces by hand a tensor factor to the space of states involving ghosts. In this subsection, we will demonstrate how this Hilbert space arises naturally in the crossed product algebra as a simple extension of $\mathcal{H}_{aux}$ introduced in Section \ref{sec: RAQ}. Harnessing the relationship between the crossed product and the extended phase space, the presence of ghost degrees of freedom come as no surprise. Indeed, one of the main motivations for constructing the extended phase space was to take advantage of the fact that ghost degrees of freedom manifest as differential forms in the vertical subbundle of an Atiyah Lie algebroid \cite{Jia:2023tki}.

In the following we will concentrate our discussion of BRST from the point of the view of the crossed product algebra $\mathcal{M} \rtimes_{\alpha} G$. We will now generalize the construction \eqref{ext0 rep} to realize the crossed product in terms of operators acting on the Hilbert space $\mathcal{H}_{ext} \equiv \Omega^{\bullet}(G;\mathcal{H}) \simeq \mathcal{H}\otimes \Omega^{\bullet}(G)$. Here $\Omega^{\bullet}(G;\mathcal{H})$ is the vector space of differential forms on $G$ taking values in $\mathcal{H}$. The aforementioned vector space is promoted to a Hilbert space by means of the following inner product:
\beq \label{ext inner}
	K(\xi,\eta) \equiv \int_G \; h\bigg(\xi \wedge \star \eta \bigg),
\eeq
where $\star$ is the Hodge dual induced by a left invariant metric on $G$ and $\xi, \eta \in \mathcal{H}_{ext}$.\footnote{The notation in \eqref{ext inner} should be read as
\beq
	h\bigg(\xi \wedge \star \eta\bigg) = h_{ab} \; \xi^a \wedge \star \eta^b,
\eeq
with $\xi = \xi^a \otimes \un{e}_a$ and $\eta = \eta^b \otimes \un{e}_b$ coordinatizations of $\xi$ and $\eta$ with respect to a basis for $\mathcal{H}$, and each $\xi^a, \eta^b \in \Omega^{\bullet}(G)$. For the metric on $G$ we take
\beq
	k = k_{AB} \; \varpi^{A} \otimes \varpi^{B},
\eeq
where $\varpi \in \Omega^1(G;\mathfrak{g})$ is the left invariant Maurer-Cartan form. In the case that $G$ is semi-simple, $G$ reduces to the Killing metric.} Notice that \eqref{ext inner} reduces to \eqref{ext0 inner} in the case that $\xi, \eta \in \Omega^0(G;\mathcal{H})$. We can similarly extend the representations \eqref{ext0 rep} as\footnote{Here
\beq
	L: G \times G \rightarrow G, \; L_t(g) = t^{-1}g
\eeq
is the left action of $G$ on itself. 
}
\beq \label{ext rep}
	\bigg( \pi_{\alpha}(x) \big(\xi\big) \bigg)(g) \equiv \pi \circ \alpha_{g^{-1}}(x)\big(\xi(g) \big), \qquad \bigg( \lambda(g')\big(\xi \big) \bigg)(g) \equiv \bigg(L_{g'}^* \xi\bigg)(g). 
\eeq
Again, \eqref{ext rep} reduces to \eqref{ext0 rep} when $\xi \in \Omega^0(G;\mathcal{H})$.\footnote{Note that the unitarity of $\lambda(t)$ is manifest via the left invariance of the metric.}

Working in the Hilbert space $\mathcal{H}_{ext}$ it is now straightforward to realize the BRST complex. To begin, we identify $T_e G \simeq \mathfrak{g}$ where $\mathfrak{g}$ is the Lie algebra integrating to $G$. Then, we can introduce the following (potentially unbounded) operators which act on $\mathcal{H}_{ext}$ through their action on the tensor factor $\Omega^{\bullet}(G)$:
\beq \label{Ghosts and Antighosts}
	b: \mathfrak{g} \rightarrow \mathcal{L}(\mathcal{H}_{ext}), \quad \un{\mu} \mapsto i_{\un{\mu}}, \qquad \un c: \mathfrak{g}^* \rightarrow \mathcal{L}(\mathcal{H}_{ext}), \quad \beta \mapsto \beta \wedge.
\eeq
Clearly, $b(\un{\mu}): \Omega^p(G;\mathcal{H}) \rightarrow \Omega^{p-1}(G;\mathcal{H})$ and $\un c(\beta): \Omega^p(G;\mathcal{H}) \rightarrow \Omega^{p+1}(G;\mathcal{H})$. 
These operators satisfy the relations
\beq \label{BRST anticomm}
	\{b(\un{\mu}), b(\un{\nu})\}  = 0= \{\un c(\beta), \un c(\gamma)\}, \quad \{\un c(\beta),b(\un{\mu})\} = i_{\un{\mu}}\beta,
\eeq
where $\{\cdot, \cdot\}$ is the anticommutator with respect to the composition of maps. Eq. \eqref{BRST anticomm} are precisely the anticommutation relations  which characterize the BRST complex. We can present these equations in a more familiar form if we introduce dual bases for $\mathfrak{g} = \text{span}\{\un{t}_A\}_{A = 1}^{\text{dim}(G)}$ and $\mathfrak{g}^* = \{t^A\}_{A = 1}^{\text{dim}(G)}$, and define the corresponding indexed operators $b_A \equiv b(\un{t}_A)$, and $c^A \equiv c(t^A)$. Then we can rewrite \eqref{BRST anticomm} as
\beq
	\{b_A,b_B\} = \{c^A,c^B\} = 0, \; \{c^A,b_B\} = \delta^A{}_B. 
\eeq

In tandem with the maps $b$ and $c$, we can also introduce a Lie algebra representation on the non-extended Hilbert space $\mathcal{H}$. To realize such a map, we regard the unitary group representation $U: G \rightarrow B(\mathcal{H})$ as arising from the exponentiation of a Lie algebra representation $v_{\mathcal{H}}: \mathfrak{g} \rightarrow \mathcal{L}(\mathcal{H})$ such that $U\big(\text{exp}(\un{\mu})\big) = \text{exp} \circ v_{\mathcal{H}}(\un{\mu})$.\footnote{By a representation, we mean that it respects the $ad$ action of $\mathfrak{g}$ on itself, $v_{\mathcal{H}}([\un\mu,\un\nu]_{\mathfrak{g}})=[v_{\mathcal{H}}(\un\mu),v_{\mathcal{H}}(\un\nu)]$ where on the right-hand side we mean the commutator of the composition of operators on $\mathcal{H}$.} We can combine the maps $v_{\mathcal{H}}$, $b$, and $c$ to form a representation $v_{ext}: \mathfrak{g} \rightarrow \mathcal{L}(\mathcal{H}_{ext})$ as
\beq \label{Extended Lie algebra Rep}
	v_{ext}(\un{\mu}) \equiv 
	 \mu^A \big(v_{\mathcal{H}}(\un{t}_A)- f_{AB}{}^C c^B b_C\big),
\eeq
where $f_{AB}{}^C$ are the structure constants of the algebra. To understand \eqref{Extended Lie algebra Rep} it is instructive to consider its action on an element $\alpha = \alpha_A \otimes t^A \in \Omega^1(G;\mathcal{H})$:
\beq \label{Extended Rep 1}
	v_{ext}(\un{\mu})(\alpha) = v_{\mathcal{H}}(\un{\mu})(\alpha_A) \otimes t^A - \alpha_C \otimes \mu^A f_{AB}{}^C t^B.
\eeq
The first term in \eqref{Extended Rep 1} corresponds to the action of the non-extended representation on the components of the form $\alpha$ which are each elements of $\mathcal{H}$. The second term in \eqref{Extended Rep 1} is precisely the co-adjoint action of $\mathfrak{g}$ on its dual $\mathfrak{g}^*$. Recall, the co-adjoint representation is a map $v_{\mathfrak{g}^*}: \mathfrak{g} \rightarrow \text{End}(\mathfrak{g}^*)$ which is defined through its duality  with the adjoint action as
\beq
	\big(v_{\mathfrak{g}^*}(\un{\mu})(\alpha)\big)(\un{\nu}) = -\alpha\big([\un{\mu},\un{\nu}]\big), \; \forall \alpha \in \mathfrak{g}^*, \un{\mu}, \un{\nu} \in \mathfrak{g}. 
\eeq
Thus, we can read \eqref{Extended Rep 1} as
\beq \label{Extended Rep 2}
	v_{ext}(\un{\mu})(\alpha) = v_{\mathcal{H}}(\un{\mu})(\alpha_A) \otimes t^A + \alpha_A \otimes v_{\mathfrak{g}^*}(\un{\mu})(t^A). 
\eeq	
Eq. \eqref{Extended Rep 2} makes clear the role of the extra terms appearing in \eqref{Extended Lie algebra Rep} relative to the Lie algebra representation $v_{\mathcal{H}}$ for the Hilbert space alone: these terms carry the action of the Lie algebra on the dual Lie algebra elements. As this action is implemented co-adjointly it is moreover clear that \eqref{Extended Lie algebra Rep} is a proper representation. On higher order forms the second term in \eqref{Extended Lie algebra Rep} will simply act via the co-adjoint action on each of the generators of the Lie algebra dual. 

To study the BRST cohomology we introduce the operator
\beq \label{BRST Operator}
	Q \equiv v_{ext}(c^A \otimes \un{t}_A) =
	c^A v_{\mathcal{H}}(\un{t}_A) - \frac{1}{2} f_{AB}{}^C c^A c^B b_C.
\eeq
The  reader may recognize \eqref{BRST Operator} as precisely the coboundary operator generating Lie algebra cohomology as discussed, for example, in \cite{Jia:2023tki,de1998lie}. As a consequence of the fact that $v_{ext}$ is a Lie algebra representation and that the $c$ operators anticommute, $Q$ is nilpotent: $Q^2 = 0$. Thus we can introduce the sets
\beq
	Z^p(G;\mathcal{H}) \equiv \{\xi \in \Omega^p(G;\mathcal{H}) \; | \; Q(\xi) = 0\}, \qquad B^p(G;\mathcal{H}) \equiv \{\xi \in \Omega^{p}(G;\mathcal{H}) \; | \; \xi = Q(\eta), \; \eta \in \Omega^{p-1}(G;\mathcal{H})\}, 
\eeq
and identify the $p^{th}$ cohomology class  with the quotient $H^p(G;\mathcal{H}) \equiv Z^p(G;\mathcal{H})/B^p(G;\mathcal{H})$. 

The significance of the cohomology classes $H^p(G;\mathcal{H})$ arise from the following observation:
\beq \label{BRST relation}
	\{Q, b(\un{\mu})\} = v_{ext}(\un{\mu}). 
\eeq
In fact, \eqref{BRST relation} is nothing but the Hilbert space version of the familiar Cartan relation between the exterior derivative, the vector contraction, and the Lie derivative:
\beq \label{Cartan Relation}
	\{d, i_{\un{X}}\} = \mathcal{L}_{\un{X}}.
\eeq
This is not an analogy: \eqref{BRST relation} is precisely equivalent to \eqref{Cartan Relation} in the context of Lie algebra cohomology with values in the representation space $\mathcal{H}$. Using \eqref{BRST relation} we can deduce that cohomology classes are preserved under the extended action of the group. For $\xi \in H^p(G;\mathcal{H})$,  under the action of $G$ generated infinitesimally by the element $\un{\mu} \in \mathfrak{g}$ we have:
\beq \label{BRST Invariance}
	\xi \mapsto \xi + v_{ext}(\un{\mu})(\xi) = \xi + Q\bigg(b(\un{\mu})(\xi)\bigg) + b(\un{\mu}) \cancelto{0}{Q(\xi)} = \xi + Q\bigg(b(\un{\mu})(\xi)\bigg) \sim \xi.
\eeq
In the first equality we have used \eqref{BRST relation}, and the final equality is up to cohomology. An alternative view point on \eqref{BRST Invariance} is that the cohomology classes of \eqref{BRST Operator} correspond to $G$-orbits in $\mathcal{H}_{ext}$. By definition, if one takes a point in $\mathcal{H}_{ext}$ and averages it over the group  an element will be obtained that remains in the same $G$-orbit. Thus, the group averaging approach discussed in Section \ref{sec: RAQ} may be regarded as a specific approach to identifying BRST cohomology classes. Both are sufficient approaches to implementing the $G$-constraints.

\subsection{Path Integral Quantization} \label{sec: Path}

As we have seen, both the Refined Algebraic and BRST quantization procedures may be interpreted as implementing constraints by imposing an equivalence relation identifying points along orbits of the action of the constraint group. In this section, we will argue that the path integral can also be viewed in this way. Our exposition is related to the usual Faddeev-Popov formalism, but leverages the geometric formulation inherent in the extended phase space to give it a new interpretation which, as we show, fits very nicely with the other quantization structures. In particular, we provide an outline of how the Faddeev-Popov procedure can be interpreted from the operator algebra perspective as a (generalized) conditional expectation related to a group averaging map. 

Again we consider a symplectic geometry $(X,\Omega)$ acted upon by a group $G$, denoting by $\mathcal{M}^{pq}_{X}$ the Poisson algebra associated with $(X,\Omega)$. For the moment, let us ignore the $G$-action on $X$ entirely. Instead, we wish to concentrate on providing an operator algebraic interpretation for the path integral. Generally speaking, a path integral can be regarded as a map, $\varphi: \mathcal{M}^{pq}_{X} \rightarrow \mathbb{C}$, which assigns to each element of the Poisson algebra a number that we interpret as its expectation value. Conventionally, we regard this mapping to be of the form:
\beq \label{Sketch of path integral}
	\varphi(f) \equiv \int \nu_{\varphi}(x) f(x(t)),
\eeq
with $x: I \subset \mathbb{R} \rightarrow X$ specifying a parameterized curve in phase space. For example, one might like to think of
\beq
	\nu_{\varphi}(x) = [dx(t)] \; e^{iS(x)},
\eeq
where $S(x)$ is a functional which one identifies with the phase space action as defined in \eqref{Definition of the action}, and
\beq
	[dx(t)] \sim \prod_{t' \in I} \text{Vol}_{X}(x(t'))
\eeq 
is the standard path integral measure on phase space. For ease of notation, we will hereafter denote \eqref{Sketch of path integral} simply by
\beq
	\varphi(f) \equiv \int_{X} \nu_{\varphi}(x) f(x),
\eeq
with the understanding that $x$ refers to a generally ``time-dependent" element of $X$, while the operator insertion is localized at points along a curve $\gamma$ in $X$. Regardless of the fine-grained form of \eqref{Sketch of path integral}, it is natural to regard $\varphi$ as defining a weight on the Poisson algebra $\mathcal{M}^{pq}_X$. We denote the normalization of \eqref{Sketch of path integral}
\beq\label{Path Integral}
Z_{\varphi} \equiv \varphi(1) = \int_X \nu_{\varphi}(x). 
\eeq 

Having established our desired interpretation for the path integral as defining a weight on the prequantum algebra of observables, let us now address the problem of implementing the constraints. In particular, we resurrect the $G$ action\footnote{Here $G$ should be interpreted as the space of curves from $I$ into the constraint group, in a notation analogous to that discussed above.} $a: G \times X \rightarrow X$. Given a point $x \in X$, the $a$-orbit of $x$ corresponds to the set of points that can be reached by $x$ upon applying the action $a$:
\beq
	a_G(x) \equiv \{x' \in X \; | \; \exists g \in G \text{ s.t. } a_g(x) = x' \}. 
\eeq
The quotient space $X/G$ can be understood as the set of such orbits.  At present we are interested in the case where $a$ preserves the measure of \eqref{Path Integral}:
\beq \label{Invariance of Omega and S}
	a_g^* \nu_{\varphi} = \nu_{\varphi}. 
\eeq
In other words, from the perspective of the path integral \eqref{Path Integral}, all points along an orbit of the action $a$ are identical.

In the usual approach, one would motivate Faddeev-Popov gauge fixing at this point by arguing that the invariance of the path integrand under the group $G$ implies that \eqref{Path Integral} overcounts the number of distinct configurations. In keeping with the theme of our operator centric approach, we take a different but ultimately equivalent perspective. Consider the following observation: generic functions in the Poisson algebra $\mathcal{M}^{pq}_{X}$ needn't share the symmetries \eqref{Invariance of Omega and S}. The weight \eqref{Sketch of path integral} does not disallow the computation of expectation values associated with these operators and thus there is an immediate risk of violating the constraints of the theory. To remedy this problem we must modify the weight $\varphi$ in such a way that it automatically enforces the symmetry constraints for any observable that we insert into it. From this point of view it is abundantly clear that the role of the Faddeev-Popov insertion in the path integral is precisely the same as the procedures outlined in Sections \ref{sec: RAQ} and \ref{sec: BRST}, namely to project generic functions down to the algebra of $G$-invariant functions. Thus the quantization procedure leads directly to properly dressed operators, from the canonical perspective.

In following with the standard Faddeev-Popov approach, the simplest way to implement the desired symmetry constraint is to construct a unique representative of each group orbit; this defines a projection from the phase space $X$ to the quotient space $X/G$:
\beq \label{Quotient Projection}
	\pi: X \rightarrow X/G, \qquad x \mapsto [x].
\eeq
Composing an arbitrary element $f \in \mathcal{M}^{pq}_X$ with the projection $\pi$ we realize a function which is, by construction, $G$-invariant:
\beq
	a_g^*\big(f \circ \pi\big)(x) = f \circ \pi \circ a_g(x) = f \circ \pi(x), \; \forall x \in X.
\eeq
In other words, the mapping
\beq \label{Gauge Fixing w projection}
	f \mapsto f \circ \pi
\eeq
takes an arbitrary element of $\mathcal{M}^{pq}_{X}$ and turns it into a gauge-fixed version  satisfying the $G$-symmetry constraint. A particularly nice way of realizing the projection \eqref{Quotient Projection} -- as utilized in the standard Faddeev-Popov procedure -- is to introduce a group non-invariant map ${\cal F}: X \rightarrow \mathfrak{g}$ whose kernel intersects each gauge orbit exactly once. That is,
\beq \label{Gauge Fixing function}
	\forall x \in X \; \exists! \; [x] \in a_G(x) \text{ s.t. } {\cal F}([x]) = 0.
\eeq
We subsequently take $[x]$ to be the representative of the orbit $a_G(x)$, thereby defining a projection \eqref{Quotient Projection}. Equivalently, \eqref{Gauge Fixing function} defines a section\footnote{There are well known topological obstructions to defining global sections in the configuration/phase space of a gauge theory as addressed, for example, in the Gribov ambiguity. These issues do not have bearing on the analysis presented here which utilizes \eqref{Gauge Fixing function} and \eqref{Def of z} solely for the purpose of constructing a conditional expectation based on a \emph{local} measure \cite{Cordes:1994fc}.} $z: X \rightarrow G$ of the extended phase space $X_{ext} \simeq X \times G$, such that
\beq \label{Def of z}
	a_{z(x)}(x) = [x], \; \forall x \in X.
\eeq

The appearance of the section $z$ provides an indication of what our next step should be -- namely we pass to $X_{ext}$, the extended phase space obtained from the symplectic manifold $(X,\Omega)$ with respect to the $G$-action defined by $a$. $X_{ext.}$ is a principal bundle but locally a point in $X_{ext}$ corresponds to the pair $(x,g)$ where $x \in X$ and $g \in G$. In this regard we can view the Poisson algebra of $X_{ext}$ as the set of maps from the group $G$ into the Poisson algebra of $X$, 
\beq \label{Extended Poisson Algebra}
	\mathcal{M}^{pq}_{X_{ext}} \equiv \Omega^0(G;\mathcal{M}^{pq}_{X}) = \{\mathfrak{F}: G \rightarrow \mathcal{M}^{pq}_{X}\},
\eeq
or what is the same, the set of functions on the product space $X \times G$. In particular, to each $f \in \mathcal{M}^{pq}_{X}$ we can associate an ``extended" function:
\beq \label{inclusion of M_0 in M_ext}
	i: \mathcal{M}^{pq}_{X} \rightarrow \mathcal{M}^{pq}_{X_{ext}}, \qquad \bigg(i(f)\bigg)(x,g) = f \circ a_g(x).
\eeq
The function $i(f)$ contains the information of $f$ evaluated along each of its gauge orbits. 

From the Poisson algebra perspective, $i$ is a morphism -- following from the fact that the action $a$ preserves the symplectic form. This means that $\{i(f),i(g)\}_{X_{ext.}}$ is $i(\{f,g\}_X)$. Equivalently the symplectic form $\Omega^{ext.}$ pulls back to $\Omega = i^*\Omega^{ext.}$, and Hamiltonian vector fields on $X_{ext.}$ are push forwards $\un V_{i(f)}=i_*\un V_f$, where $\un V_f$ is a Hamiltonian vector field on $X$. These observations fit into the larger view of $\mathcal{M}^{pq}_{X_{ext}}$ as a ``pre"-crossed product algebra comprised of two kinds of elements. First, there are the extended functions comprising the image of \eqref{inclusion of M_0 in M_ext} which furnish a copy of the non-extended Poisson algebra $\mathcal{M}^{pq}_{X}$. Second, there are functions representing each infinitesimal generator of $G$, furnishing a copy of its Lie algebra. The complete Poisson algebra $\mathcal{M}^{pq}_{X_{ext}}$ therefore contains a copy of the non-extended algebra, a copy of the Lie algebra of $G$, and cross terms associated with the infinitesimal action of $G$ on $\mathcal{M}^{pq}_{X}$. For a more formal derivation of this algebra, see \cite{Klinger:2023tgi}. 

Let us now define the Faddeev-Popov determinant:
\beq \label{FP Det}
	\Delta_{{\cal F}}(x)^{-1} \equiv \int_{G} \mu(g) \; \delta({\cal F}\circ a_g(x)).
\eeq
Here $\mu$ is the left invariant Haar measure of the group, and ${\cal F}$ is the gauge fixing functional.\footnote{See Appendix \ref{app: FP} for a review of how the Faddeev-Popov determinant can be evaluated in terms of a Berezin integral over a pair of ghost variables $c$ and $\overline{c}$.} It is straightforward to show that \eqref{FP Det} is invariant under the action of $G$:
\begin{flalign}
	\Delta_{{\cal F}}\circ a_{h}(x)^{-1} &= \int_{G} \mu(g) \; \delta({\cal F}\circ a_{hg}(x)) \nonumber \\
	&= \int_{G} \mu(h^{-1}k) \; \delta({\cal F}\circ a_{k}(x)) \nonumber \\
	&= \int_{G} \mu(k) \; \delta({\cal F}\circ a_{k}(x)) = \Delta_{{\cal F}}(x)^{-1}.
\end{flalign}
Together, the FP determinant and the delta function define a conditional probability distribution
\beq \label{definition of Thom}
	\Phi(g \mid x) = \Delta_{{\cal F}}\circ a_g(x) \; \delta({\cal F}\circ a_g(x))
\eeq
characterized by the following properties:
\begin{flalign} \label{Thom Cond Exp}
	&\int_{G} \mu(g) \; \Phi(g \mid x) = 1, \; \forall x \in X \nonumber \\
	&\int_{G} \mu(g) \; \Phi(g \mid x) i(f)(x,g) = f([x]). 
\end{flalign}
Here we have used the fact that for each $x$ the unique element $g$ that solves ${\cal F}\circ a_g(x) = 0$ is $g = z(x)$, and $a_{z(x)}(x) = [x]$. By extension, we can use $\Phi(g \mid x)$ to define a conditional expectation 
\beq \label{cond exp}
	T: \mathcal{M}^{pq}_{X_{ext}} \rightarrow \mathcal{M}^{pq}_{X}, \qquad \mathfrak{F}(x,g) \mapsto \int_{G} \mu(g) \; \Phi(g \mid x) \mathfrak{F}(x,g) = \mathfrak{F}(x,z(x)). 
\eeq
As is evident from \eqref{Thom Cond Exp} the composition of \eqref{inclusion of M_0 in M_ext} and \eqref{cond exp} turns out to be precisely the gauge fixing map we are looking for:
\beq \label{Cond exp of M_0}
	T \circ i(f)(x,g) = f([x]). 
\eeq 

We are now prepared to define an extended path integral construction which properly accounts for the group of constraints. In particular, we define a ``dual weight" $\tilde{\varphi}: \mathcal{M}^{pq}_{X_{ext}} \rightarrow \mathbb{C}$ given explicitly by
\beq \label{Path Integral Dual Weight}
	\tilde{\varphi} \equiv \varphi \circ T.
\eeq
Acting on an element of the non-extended Poisson algebra we have\footnote{Here we have used the fact that $\nu_{\varphi}$ is presumed to be invariant under the group action.}
\beq \label{Gauge Fixing}
	\tilde{\varphi} \circ i(f) = \int_{X_{ext}} \nu_{\varphi}^{ext}(x,g) \; i(f) = \int_{X_{ext}} \nu_{\varphi}(x) \wedge \mu(g)\; \Phi(g \mid x) \; f \circ a_g(x) = \int_{X/G} \nu_{\varphi}^{X/G}([x]) \; f([x]). 
\eeq
Here $\nu_{\varphi}^{ext}$ is the path integral measure obtained by promoting $\nu_{\varphi}$ to the extended phase space. The second equality in \eqref{Gauge Fixing} is obtained by first performing the integration over the group to activate the gauge fixing, with $\nu_{\varphi}^{X/G}$ the path integral measure on the quotient space obtained from pushing forward $\nu_{\varphi}^{ext}$ by the projection \eqref{Quotient Projection} in the measure theoretic sense -- including a factor of the FP determinant. 

At this point we should pause to make an observation. The constraint quantization defined by \eqref{Gauge Fixing} is distinct from the usual Faddeev-Popov formalism in the following way -- rather than mandating that the insertion $f$ be invariant we have allowed for the insertion of arbitrary ``dressed" functions, and then used the fact that \eqref{Cond exp of M_0} renders such elements $G$-equivariant. In this respect it is not true that the expectation value of a non-invariant insertion is zero in \eqref{Gauge Fixing function}. Rather, \eqref{Gauge Fixing} computes the expectation value of the gauge-invariant projection of any given insertion. Here the gauge invariant projection of a generic operator can be identified with the gauge orbit to which it belongs. These ideas will become more manifest in the examples discussed in Section \ref{sec: Zamples}.

It is straightforward to show that the new state obtained by promoting $\varphi \mapsto \tilde{\varphi} \circ i$ is invariant under the action of the group. In particular,
\begin{flalign} \label{Invariance of promoted state}
	\tilde{\varphi} \circ i(f \circ a_{g'}) &= \int_{X_{ext}} \nu_{\varphi}(x) \wedge \mu(g) \; \Phi(g \mid a_{g'}(x)) f \circ a_{g}(a_{g'}(x)) \nonumber \\
	&= \int_{X_{ext}} \nu_{\varphi}(y) \wedge \mu(g) \; \Phi(g \mid y) f \circ a_{g}(y) = \tilde{\varphi} \circ i(f), \; \forall g' \in G.
\end{flalign}
To move from the first to the second line we have changed variables, $y = a_{g'}(x)$, and used \eqref{Invariance of Omega and S} to change the measure $\nu_{\varphi}$ with impunity. Recognizing $f \circ a_{g'} = a_{g'}^*f$ and using the definition of the Lie derivative, \eqref{Invariance of promoted state} implies:
\beq \label{Invariance of promoted state 2}
	\tilde{\varphi} \circ i(\mathcal{L}_{\xi_{\un{\mu}}} f) = 0,
\eeq 
where $\xi_{\un{\mu}}$ is the tangent vector generating the action $a_{g'}: X \rightarrow X$ as its integral curves, with $\un{\mu} \in \mathfrak{g}$ the Lie algebra element integrating to $g' \in G$. As discussed in \cite{Klinger:2023tgi}, in the extended phase space $\mathcal{L}_{\xi_{\un{\mu}}} f = \{H_{\un{\mu}},f\}_{ext.}$, where $H_{\un{\mu}}$ is the Hamiltonian function associated with the group action generated by $\un{\mu}$ and $\{,\}_{ext.}$ is the Poisson bracket. Thus, \eqref{Invariance of promoted state 2} further implies that
\beq \label{Invariance of promoted state 3}
	\tilde{\varphi} \circ i(\{H_{\un{\mu}},f\}_{ext.}) = 0.
\eeq
Eqn. \eqref{Invariance of promoted state 3} indicates that the Hamiltonian functions $H_{\un{\mu}}$ act trivially inside of expectation values for physical states.

\subsubsection{Dual Weights and Constraint Quantization}

Having outlined how the Faddeev-Popov gauge fixing procedure works in the path integral sense, let us conclude this section by discussing how this approach can be connected to the crossed product $\mathcal{M} \rtimes_{\alpha} G$ via the dual weight theorem. To make this comparison we employ the correspondence \eqref{commutative diagram}. Let $\mathcal{M}$ be a von Neumann algebra acted upon by the group $G$ via the automorphism $\alpha: G \times \mathcal{M} \rightarrow \mathcal{M}$. One should view $\mathcal{M}$ as the quantized version of the non-extended Poisson algebra $\mathcal{M}^{pq}_{X}$. We denote by $\mathcal{M}_{ext}$ the set of strongly continuous, compactly supported maps from the group $G$ into the von Neumann algebra $\mathcal{M}$. This is inspired by the definition \eqref{Extended Poisson Algebra} for the Poisson algebra of the extended phase space. As we shall see, the von Neumann algebra associated with $\mathcal{M}_{ext}$ is equivalent to the crossed product $\mathcal{M} \rtimes_{\alpha} G$, and may therefore be regarded as the quantization of $\mathcal{M}^{pq}_{X_{ext}}$. We will denote elements of $\mathcal{M}_{ext}$ by $\mathfrak{X},\mathfrak{Y},$ etc. with the understanding that these objects are maps i.e., $\mathfrak{X}(g) \in \mathcal{M}$ for each $g \in G$.

To explicitly study $\mathcal{M}_{ext}$ as a $C^*$ algebra, and to formulate its associated von Neumann algebra we follow the construction of Haagerup \cite{Haagerup1978I, hiai2020concise}. Recall that integration on the group $G$ is defined in terms of a left invariant Haar measure, $\mu$, along with its module function\footnote{The module function is a group homomorphism, $\delta(g)\delta(h) = \delta(gh)$, defined by the property
\beq
	\int_G \mu(g) f(g) = \delta(h) \int_G \mu(g) f(gh).
\eeq
} $\delta: G \rightarrow \mathbb{C}$ which tracks the failure of $\mu$ to be right invariant. The set $\mathcal{M}_{ext}$ can be made into an involutive Banach algebra by endowing it with the following operations which define a product and an involution, respectively:
\beq \label{Product on K}
	\bigg(\mathfrak{X} \boldsymbol{\cdot} \mathfrak{Y}\bigg)(g) \equiv \int_G \mu(h) \; \alpha_h\bigg(\mathfrak{X}(gh)\bigg) \mathfrak{D}(h^{-1}),
\eeq
\beq \label{Involution on K}
	\mathfrak{X}^{\star}(g) \equiv \delta(g^{-1}) \alpha_{g^{-1}}\bigg(\mathfrak{X}(g^{-1})\bigg)^*.
\eeq
Given the representation \eqref{ext0 rep}, we can define a $\star$-representation of $\mathcal{M}_{ext}$ acting\footnote{Again, this result can be generalized to any covariant representation.} on $\mathcal{H}_{aux}$:
\beq \label{Rep on C* CP}
	\rho: \mathcal{M}_{ext} \rightarrow B(\mathcal{H}_{aux}), \; \rho(\mathfrak{X}) \equiv \int_G \mu(g) \; U(g) \pi\bigg(\mathfrak{X}(g)\bigg). 
\eeq
In \cite{Haagerup1978I} it is shown that $\text{im}(\rho)$ is dense in $\mathcal{M} \rtimes_{\alpha} G$ in the weak operator topology. Thus, the weak closure of $\mathcal{M}_{ext}$ is equivalent to the crossed product algebra. 

We now come to the crucial topic of dual weights. To begin, we define the map:
\beq \label{Dual weight CE}
	T: \mathcal{M}_{ext} \rightarrow \mathcal{M}, \; \mathfrak{X} \mapsto \mathfrak{X}(e).
\eeq
Here $e \in G$ is the identity element. In other words, \eqref{Dual weight CE} is nothing other than a projection of $\mathcal{M}_{ext}$ down to $\mathcal{M}$, as in \eqref{Thom Cond Exp}. As demonstrated by Haagerup in \cite{Haagerup1978II,Haagerup1978I}, this projection can be used to define an operator-valued weight from the crossed product to the algebra $\mathcal{M}$. A faithful, semi-finite, normal weight $\varphi$ on $\mathcal{M}$ corresponds to a quantum state, and therefore plays the same role as the path integral \eqref{Sketch of path integral}. Combining such a weight with the map \eqref{Dual weight CE} we obtain a weight on the extended algebra:
\beq \label{T Dual Weight}
	\tilde{\varphi} \equiv \varphi \circ T.
\eeq
Of course, this is totally analogous to \eqref{Path Integral Dual Weight}.

We would now like to make use of \eqref{T Dual Weight} in order to construct $G$-invariant weights on $\mathcal{M}$. To do so, we take motivation from \eqref{inclusion of M_0 in M_ext} and specify an explicit embedding of the algebra $\mathcal{M}$ inside of $\mathcal{M}_{ext}$. Since $\mathcal{M}$ is linearly spanned by the set of positive elements therein, we define this procedure on positive operators $x^*x \in \mathcal{M}$ and then extend the definition appropriately.\footnote{The purpose of defining \eqref{Embedding} in such an apparently circuitous way is to ensure that positive operators in $\mathcal{M}$ remain positive with respect to the involution and product structure defining $\mathcal{M}_{ext}$.} The relevant embedding map is of the form:
\beq \label{Embedding}
	i: \mathcal{M} \rightarrow \mathcal{M}_{ext}, \; x^* x \mapsto \alpha(x)^{\star} \boldsymbol{\cdot} \alpha(x).
\eeq
In \eqref{Embedding} the notation $\alpha(a)$ corresponds to the element $\alpha(a): G \rightarrow \mathcal{M}$ in $\mathcal{M}_{ext}$ such that $\bigg(\alpha(a)\bigg)(g) = \alpha_g(a)$. Using \eqref{Product on K} and \eqref{Involution on K} it is easy to compute:\footnote{We have also used the fact that $\alpha$ is a $*$-homomorphism: $\alpha_g(x^*y) = \alpha_g(x^*)\alpha_g(y)$.}
\begin{flalign} \label{Gauge fixing = Group Averaging}
	T\circ i(x^*x) = \bigg(\alpha(x)^{\star} \boldsymbol{\cdot} \alpha(x)\bigg)(e) &= \int_G \mu(g) \; \alpha_g(x)^* \alpha_g(x) \\
	&= \int_G \mu(g) \; \alpha_g(x^* x) = \Pi_d(x^* x),
\end{flalign}
which is precisely \eqref{Group averaging in algebra} corresponding to the dressing of the operator $x$. Thus, as was the case in \eqref{Cond exp of M_0}, composing the maps $T$ and $i$ results in our desired mapping from $\mathcal{M}$ to the invariant subalgebra $\mathcal{M}_{inv}$. Notice that in this case rather than realizing a gauge fixing condition, \eqref{Gauge fixing = Group Averaging} results in the group averaging map from Section \ref{sec: RAQ}. In conclusion, we have shown that
\beq \label{G invariant Dual Weight}
	\tilde{\varphi} \circ i(x^*x) = \varphi \circ \Pi_d(x^*x).
\eeq

Hopefully the analogy between this analysis and the path integral analysis is clear. The algebra $\mathcal{M}_{ext}$ plays the role of the extended Poisson algebra. It is explicitly obtained by considering maps from the group $G$ into the algebra $\mathcal{M}$, as the algebra $\mathcal{M}^{pq}_{X_{ext}}$ can be thought of as the set of maps from $G$ into $\mathcal{M}^{pq}_{X}$ \eqref{Extended Poisson Algebra}. The projection \eqref{Dual weight CE} is analogous to the conditional expectation defined by the Faddeev-Popov procedure \eqref{Thom Cond Exp}, which localizes maps from $G$ down to the zero section of an appropriate fibration. In both cases, the composition of $T$ with a weight $\varphi$ of the non-extended algebra results in a dual weight, \eqref{Path Integral Dual Weight} and \eqref{T Dual Weight}, respectively. To compute the value of the dual weight applied to elements in the non-extended algebra we need a lifting map that embeds the non-extended algebra in the extended one. In the path integral context this map is supplied by \eqref{inclusion of M_0 in M_ext}, while in the algebraic context this map is supplied by \eqref{Embedding}. In both cases these maps can be interpreted as dressing the non-extended operators by acting on them with the relevant group automorphism. The projection map and the embedding then conspire to perform the gauge fixing, as can be seen in \eqref{Cond exp of M_0} and \eqref{Gauge fixing = Group Averaging}. Thus, when applied to these dressed operators, the aforementioned dual weights are manifestly $G$-invariant -- \eqref{Gauge Fixing} and \eqref{G invariant Dual Weight}. A summary of this correspondence can be found in Table \ref{Overview of Path Integral}.

\begin{table}[H] 
\centering
\begin{tabular}{|c c c|} 
\hline
& Path Integral & Crossed Product \\
\hline \hline
Non-extended Algebra: & $\mathcal{M}^{pq}_{X}$ & $\mathcal{M}$ \\
\hline
Extended Algebra: & $\mathcal{M}^{pq}_{X_{ext}} \sim \Omega^0(G;\mathcal{M}^{pq}_{X})$ & $\mathcal{M}_{ext} \sim \Omega^0(G;\mathcal{M})$ \\
\hline
Projection Map: & $T: \mathcal{M}^{pq}_{X_{ext}} \rightarrow \mathcal{M}^{pq}_{X}, \; \mathfrak{F}(x,g) \mapsto \mathfrak{F}(x,z(x))$ & $T: \mathcal{M}_{ext} \rightarrow \mathcal{M}, \; \mathfrak{X}(g) \mapsto \mathfrak{X}(e)$ \\
\hline
Inclusion: & $i: \mathcal{M}^{pq}_{X} \rightarrow \mathcal{M}^{pq}_{X_{ext}}, \; f(x) \mapsto f \circ a_g(x)$ & $i: \mathcal{M} \rightarrow \mathcal{M}_{ext}, \; x^*x \mapsto \alpha(x)^{\star} \boldsymbol{\cdot} \alpha(x)$ \\
\hline
Implementing Constraint: & $T \circ i(f) = f \circ \pi$ & $T \circ i(x) = \Pi_d(x)$ \\
\hline
Dual Weights: & $\tilde{\varphi} = \varphi \circ T$ & $\tilde{\varphi} = \varphi \circ T$ \\
\hline
Invariant States: & $\tilde{\varphi} \circ i = \varphi \circ \pi^*$ & $\tilde{\varphi} \circ i = \varphi \circ \Pi_d$ \\
\hline
\end{tabular}
\caption{Overview of the correspondence between the path integral approach to constraint quantization, and the same approach using the crossed product and dual weights. This can be regarded as an extension of the correspondence \eqref{commutative diagram}.}
\label{Overview of Path Integral}
\end{table}

\subsection{Commutation Theorem for Crossed Products} \label{sec: comm thm}

This section details an alternative approach to implementing constraints via the crossed product algebra by augmenting the action of the gauge group. As we shall see, this approach is equivalent to the constraint quantization procedures we have introduced, and fits naturally into the extended phase space formalism. The discussion in this section may be regarded as further explicating the connection between the conditional expectation oriented approach to implementing constraints, and the related approach of designating a so-called observer. 

Let $({\cal M},G,\alpha)$ denote a covariant system; that is ${\cal M}$ is a von Neumann algebra acted upon via the $G$-automorphism $\alpha: G \times {\cal M} \rightarrow {\cal M}$. Let $\pi: {\cal M} \rightarrow B({\cal H})$ be a Hilbert space representation of ${\cal M}$, and denote by $\mathcal{H}_{aux} \equiv L^2({\cal H};G,\mu) \simeq {\cal H} \otimes L^2(G,\mu)$ the canonical covariant representation space of the system $({\cal M},G,\alpha)$. As always, $\mu$ is a left invariant Haar measure on $G$. The covariant representation of $({\cal M},G,\alpha)$ on $\mathcal{H}_{aux}$ is specified in terms of the pair of maps
\beq \label{Covariant Reps}
	\pi_{\alpha}: {\cal M} \rightarrow B(\mathcal{H}_{aux}), \qquad \lambda: G \rightarrow U(\mathcal{H}_{aux}),
\eeq
where here
\beq
	\bigg(\pi_{\alpha}(x) \xi \bigg)(g) \equiv \pi \circ \alpha_{g^{-1}}(x)\big(\xi(g)\big),
\eeq
and
\beq
	\bigg(\lambda(g)\xi\bigg)(g') \equiv \xi(g^{-1} g')
\eeq
is left translation. The crossed product algebra ${\cal M} \rtimes_{\alpha} G$ is the von Neumann algebra generated by $\pi_{\alpha}(x)$ and $\lambda(g)$ closed in the weak operator topology induced by $\mathcal{H}_{aux}$.

It can be shown that $\mathcal{M} \rtimes_{\alpha} G$ is realized as a subalgebra of $\mathcal{M} \otimes B(L^2(G;\mu))$.\footnote{The proof simply amounts to showing that $\pi_{\alpha}(x)$ and $\lambda(g)$ commute with $\pi({\cal M})' \otimes \mathbb{1}$.} From this point of view, we can now state the commutation theorem, which may be taken as an alternative definition of the crossed product algebra. Let\footnote{Often $\theta$ is written as $\theta = \alpha \otimes \text{Ad}_r$. Our notation may be interpreted as $\theta = (\alpha \otimes \mathbb{1}) \circ (\mathbb{1} \otimes \text{Ad}_r)$.} $\theta \equiv \alpha \circ \text{Ad}_{r}: G \times {\cal M} \otimes B(L^2(G;\mu)) \rightarrow {\cal M} \otimes B(L^2(G;\mu))$ be an automorphism, with $r: G \rightarrow U(L^2(G;\mu))$ the right translation
\beq
	\bigg(r(g) \xi \bigg)(g') \equiv \delta(g)^{1/2} \xi(g'g). 
\eeq 
Then, ${\cal M} \rtimes_{\alpha} G$ is equivalent to the invariant subalgebra of ${\cal M} \otimes B(L^2(G;\mu))$ with respect to the action $\theta$:
\beq
	\mathcal{M} \rtimes_{\alpha} G = \bigg(\mathcal{M} \otimes B(L^2(G;\mu))\bigg)^{\theta}_{inv.}.
\eeq
For a proof of this theorem, see \cite{van1978continuous}. In fact, the proof proceeds by a group averaging argument which is very similar to the one outlined in Section \ref{sec: RAQ}.

Notice that the commutation theorem does not imply that the crossed product algebra is invariant under the action of the original automorphism $\alpha$. Instead, the invariance of the algebra has been realized in this case by augmenting the action $\alpha \mapsto \theta$. In the literature, e.g. \cite{Witten:2021unn, Chandrasekaran:2022eqq, Chandrasekaran:2022cip,Jensen:2023yxy}, this augmentation is achieved through the introduction of an `observer' whose Hamiltonian implements the right translation. In the current note, we have avoided appealing to the notion of an observer in favor of a geometric presentation of the crossed product. By consequence, we can now see exactly how these two approaches are related. In our presentation, the role of the right action is played by the conditional expectation. This is especially clear in Section \ref{sec: Path}, in which the conditional expectation projects points along generic gauge orbits to a reference point singled out by the choice of gauge fixing functional. This point will also be addressed in Section \ref{sec: Zamples}, whereupon it is observed that, after dressing the fields, a gauge transformation can be absorbed entirely into a right translation of the group element $g$ introduced into the extended phase space. 

At the same time, we can provide a compatible interpretation for the extension of the action $\alpha \mapsto \theta = \alpha \circ \text{Ad}_{\rho}$ through the symplectic analysis of the extended phase space. Let $(X,\Omega)$ be a symplectic manifold with $G$-action $a: G \times X \rightarrow X$ quantizing to the von Neumann algebra ${\cal M}$ with $G$-automorphism $\alpha: G \times {\cal M} \rightarrow {\cal M}$. The action $a$ is generated by symplectic (but not necessarily Hamiltonian) vector fields $\xi_{\un{\mu}} \in TX$, such that 
\beq
	\xi_{\un{\mu}} \equiv (a_{\text{exp}(t\un{\mu})})_* \frac{d}{dt}. 
\eeq
That is, for each Lie algebra element $\un{\mu} \in \mathfrak{g}$, we have a vector field $\xi_{\un{\mu}}$ whose integral curves generate the action $a$. When we move from $X$ to the extended phase space $X \rightarrow X_{ext}$ the action $a$ is promoted to an action $\Phi$ which is automatically Hamiltonian (and moreover equivariant). Infinitesimally, this is accomplished by mapping
\beq \label{Hamiltonian Vector Fields in Xext}
	\xi_{\un{\mu}} \mapsto -\xi_{\un{\mu}} \oplus \un{\mu} \in TX_{ext},
\eeq
where we have regarded $TX_{ext} \simeq TX \oplus \mathfrak{g}$.\footnote{More rigorously, this is done by passing to the Atiyah Lie algebroid associated with the principal bundle $X_{ext}$. See Section 4 of \cite{Klinger:2023qna} for a detailed analysis.} Eqn. \eqref{Hamiltonian Vector Fields in Xext} is the symplectic analog of the algebraic promotion $\alpha \mapsto \alpha \circ \text{Ad}_{\rho}$. 

In fact, \eqref{Hamiltonian Vector Fields in Xext} generate the right action
\beq \label{Full Right action EPS}
	R: G \times X_{ext} \rightarrow X_{ext}, \qquad R_h(g,x) = (gh, a_{h^{-1}}(x)),
\eeq
which is implicated in the structure of the extended phase space viewed as a principal $G$-bundle over $X$ \cite{Klinger:2023tgi}. Recalling the map \eqref{inclusion of M_0 in M_ext} we can now show that dressed observables $i(f) \in \mathcal{M}^{pq}_{X_{ext}}$ are automatically invariant under the extended action:
\beq \label{Invariance via commutation theorem}
	\bigg(R_h^* i(f)\bigg)(g,x) = \bigg(i(f)\bigg)(gh,a_h^{-1}(x)) = f \circ a_{gh}(a_{h^{-1}}(x)) = \bigg(i(f)\bigg)(g,x). 
\eeq
This is the symplectic precursor to the observation of the commutation theorem, namely that dressed operators $\pi_{\alpha}(x) \in \mathcal{M} \rtimes_{\alpha} G$ are invariant under the extended action $\theta$. It is also immediately clear that observables obtained via \eqref{Cond exp of M_0} will be invariant under the extended action:
\beq \label{Dressing}
	R_h^*\bigg(T \circ i(f)\bigg)(x) = a_g^*\bigg(T \circ i(f)\bigg)(x) = f \circ a_{z \circ a_h(x)}(a_h(x)) = T \circ i(f)(x) = f([x]).
\eeq
Here, we have made the identification $\mathcal{M}^{pq}_{X} \subset \mathcal{M}^{pq}_{X_{ext}}$ and used the fact that
\beq \label{GF}
	a_{z \circ a_g(x)}(a_g(x)) = a_{z(x)}(x), \; \forall g \in G, x \in X. 
\eeq
Thus, we have now demonstrated that dressing \eqref{Invariance via commutation theorem} and gauge fixing \eqref{Dressing} provide two different but compatible approaches to implementing constraints in the context of the extended phase space/crossed product algebra. 

In summary, one interpretation for the projection map that implements constraints is as a right action which compensates the gauge transformation $\alpha$ thereby rendering generic observables gauge invariant. Denoting the projection by $E$, one may therefore regard $\alpha' \equiv \alpha \circ E$ as a modified action to be compared with $\theta = \alpha \circ \text{Ad}_r$. In other words, the projection should be viewed as generating an `emergent observer', or, equivalently, the observer viewed as introducing a projection. This point of view is consistent with the correspondence between the extended phase space and the crossed product, and provides additional support to the claim that the extension provides a geometrization of the observer. We note in passing that this observation bears a close resemblance to Relational Quantum Dynamics and the Page-Wootters formalism. We refer the reader to \cite{Hoehn:2019fsy} for a recent work which provides a good entrance point to these fields.

\section{Examples} \label{sec: Zamples}

In this section we work through a series of three examples to draw out relevant features of the general theory presented in Sections \ref{sec: preliminaries} and \ref{sec: con-quant}. The examples are presented in order of increasing sophistication, culminating in a discussion of gravitational theories quantized on null hypersurfaces, in relation to the recent work \cite{Ciambelli:2023mir}. 

\subsection{The Symmetric Harmonic Oscillator}\label{sec: Oscillator}

As a simple first example, we consider gauging in a quantum mechanical system, the $D$-dimensional symmetric harmonic oscillator. Since this is a quantum mechanics system, it can be regarded as a field theory in $(0+1)$ spacetime dimensions, and consequently has some simplicity compared to the general case (for example, there are no codimension-2 structures). The phase space of the ungauged parent system is coordinatized by $\vec q,\vec p$, the Hamiltonian generating time translations is 
\beq
H_{cl.}=\frac12 (\vec{p}\,{}^2+\vec q\,{}^2),
\eeq
and we take the symplectic form to be $\Omega = \delta p_i\wedge \delta q^i$. It is well-known that this system possesses Hamiltonian vector fields generating $SU(D)$, and the states at the $n^{th}$ energy level come in the $n$-index symmetric tensor representation, for $n=0,1,...$. The operators generating $SU(D)$ are linear combinations of $\hat a^\dagger_i\hat a^j$, where $\hat a_j^\dagger=\frac12(\hat p_j+i\hat q^j)$ are a set of ladder operators. The $SO(D)\subset SU(D)$ symmetry acts as
\beq
q^j\mapsto R^j{}_kq^k,\qquad p_j\mapsto p_kR^k{}_j
\eeq
and is interpreted as spatial rotation.

In this section we consider gauging the $SO(D)$ subgroup. At the level of the Hilbert space it is clear that this requires the $SO(D)$ generators to annihilate physical states. Since the states of the ungauged theory come in particular irreps of $SU(D)$, it is a simple matter to extract the states of the gauged theory as they correspond to singlets that may occur in the decomposition of the $SU(D)$ irreps into $SO(D)$ representations. Such singlets occur for even $n$, as the trace is the invariant. Thus all states of the gauged theory are generated from the vacuum by $\hat A^\dagger:=\frac12\sum_j\hat a^\dagger_j\hat a^\dagger_j$ which indeed is an $SO(D)$-invariant.

From the operator algebra point of view, the ungauged theory has operators generated by $\hat a^\dagger_j,\hat a^j$, forming the algebra $\mathcal{M}=\CC[\hat a^\dagger_j,\hat a^j]/{\sim}$ where the equivalence refers to $[\hat a^i,\hat a^\dagger_j]=\delta^i{}_j\hat I$. This is a von Neumann algebra, with Hilbert space representation on, for example, $L^2(\RR^D)$. Gauging the $SO(D)$ symmetry in the manner outlined above yields the von Neumann algebra $\mathcal{M}_{inv.}=\mathfrak{A}(\{\hat I,\hat H,\hat A^\dagger,\hat A\})$.\footnote{Here, $\mathfrak{A}(\cdot)$ denotes the von Neumann algebra generated by the indicated elements.}
In particular, there is a (spectrum generating) $SL(2)$ algebra
\beq \label{Harmonic Gauge Invariant Algebra}
[\hat A^\dagger,\hat A]=-2\hat H,\qquad [\hat H,\hat A^\dagger]=A^\dagger,\qquad [\hat H,\hat A]=-\hat A
\eeq
where $\hat H=\frac12\sum_j\hat a^\dagger_j\hat a^j+\tfrac{D}{4}\hat I$. The spectrum consists of a single unitary lowest weight representation beginning with energy $E_0=D/4$. 

We now consider the path integral, along with a Faddeev-Popov insertion which we would like to interpret as a conditional expectation. 
Starting from the phase space $X=\{\vec q,\vec p\}$, the path integral of the ungauged theory uses Hamilton's principle function evaluated along a curve $\gamma$ in phase space
\beq
S=\int dt'\Big( \vec p(t')\cdot\dot{\vec q}(t')- \tfrac12\vec p(t')^2-\tfrac12\vec q(t')^2\Big)
\eeq
The path integral is then of the form
\beq \label{Harmonic Path Integral}
\varphi_t(f)=\int [dq(t')dp(t')]_t e^{iS}f(\vec q,\vec p)
\eeq
where $t$ makes reference to the intrinsic length of a path in phase space. Here $f$ is a compact notation for a possible series of insertions along the time contour. For simplicity, we hereafter restrict our attention to the $D=2$ case, as it already possesses the features that we wish to display. The $SO(2)$ symmetry is generated on phase space by the vector field
\beq
\un V_{\alpha}=\alpha x^j\varepsilon^{i}{}_{j}\frac{\delta}{\delta x^i}-\alpha p_i\varepsilon^{i}{}_{j}\frac{\delta}{\delta p_j}.
\eeq
 Gauging the $SO(2)\subset SU(2)$ symmetry corresponds to introducing an $SO(2)$ connection on the phase space $X$ that promotes the symmetry to be local in time. That is, we build on $X$ a principal $SO(2)$ bundle with connection, which is $X_{ext.}$. An $SO(2)$ connection includes two pieces. The first is the familiar gauge field which covariantizes time derivatives,
\beq \label{Harmonic action w constraint}
S\to \int dt\Big( \vec p\cdot D_t{\vec q}- \tfrac12\vec p^2-\tfrac12\vec q^2\Big)=S-\int dt\; A_0 L_z
\eeq
with $D_t\vec q=\pa_t\vec q-A_0 \uuline{\varepsilon}\cdot\vec q$ and thus $L_z=\vec p\cdot\uuline{\varepsilon}\cdot\vec q$.  We therefore recognize $L_z$ as the Noether current for $SO(2)$ and, since we are in $(0+1)$-dimensions, this coincides with the gauge constraint.

The second piece of the $SO(2)$ connection is a vertical component which can be thought of as $\chi=\delta\phi$, for $\phi$ an element of the Lie algebra $\mathfrak{so}(2)$. The extended phase space is coordinatized locally by $X_{ext.}=\{\vec q,\vec p,\phi\}$ with $\phi$ thought of  as a local fibre coordinate (that is, an exponential coordinate for the group $SO(2)$). The group action is given by $a: X\times G\to X$, where $a:(\vec{\tilde q},\vec{\tilde p},\phi)\mapsto a_\phi(\vec{\tilde q},\vec{\tilde p})\equiv(\vec{ q},\vec{ p})$. More specifically, $a_\phi$ is of the form  
\beq\label{rotateSHO}
\begin{pmatrix} x\cr y\end{pmatrix} \equiv \uuline{R}(\phi)\begin{pmatrix}\tilde x\cr\tilde  y\end{pmatrix},\qquad 
\begin{pmatrix} p_x\cr p_y\end{pmatrix}\equiv \uuline{R}(\phi)\begin{pmatrix}\tilde p_x\cr \tilde p_y\end{pmatrix}
\eeq
with $\uuline{R}(\phi)\in SO(2)$ a rotation through angle $\phi$. When the extended phase space is pulled back to a bundle over a curve parameterized by $t$, we should in addition include $A_0=\tilde A_0+\dot\phi$ in order to ensure the covariance of \eqref{Harmonic action w constraint}. 

We interpret \eqref{rotateSHO} as defining tilded variables, $(\vec{\tilde q}, \vec{\tilde p})$, which are invariant under the $SO(2)$ action in the sense that a transformation of the untilded variables can be absorbed into a shift of the angle $\phi$, see \eqref{rotateSHO}. In this regard $(\vec{\tilde q}, \vec{\tilde p})$ represent a single point on a gauge orbit.
We can extend this statement to an arbitrary function $f(\vec q, \vec p)$ on $X$ by introducing the map \eqref{inclusion of M_0 in M_ext} which passes $f$ to a function on the extended phase space
\beq
	\bigg(i(f)\bigg)(\vec{\tilde q},\vec{\tilde p}, \phi) = f \circ a_{\phi}(\vec{\tilde q},\vec{\tilde p})=f(\vec q,\vec p). 
\eeq 

Similarly, using \eqref{rotateSHO} the symplectic potential can be rewritten
\beqn
\theta(\vec q,\vec p)
&=&\tilde p_x\delta\tilde x+\tilde p_y\delta\tilde y-(\tilde x\tilde p_y-\tilde y\tilde p_x)\delta\phi
\\
&=&\theta(\vec {\tilde q},\vec{\tilde p})
-\tilde L_z\delta\phi\label{SHOtransformtheta}
\eeqn
The term $\delta\phi$ is the Maurer-Cartan form on the extended space of fields. This perspective is consistent with the aforementioned observation that the pair $(A_0,\phi)$ specifies a full connection on the extended phase space regarded as a principal bundle over $X$, with $\chi$ being the vertical part. As was observed in Section \ref{sec: preliminaries}, the Maurer-Cartan form and the constraint $\tilde L_z$ form a symplectic pair. 

From the algebroid perspective, what we have done in \eqref{SHOtransformtheta} is to extract a vertical, or group variational term as was introduced in \eqref{G variational term}. For the associated transformation of the symplectic structure to be canonical the count of degrees of freedom should remain the same, and thus we should project $\theta(\vec {\tilde q},\vec{\tilde p})$ to a form on $X/G$. This is analogous to introducing a choice of gauge fixing, corresponding to the addition into the path integral of a conditional probability measure:
\beq \label{Thom Harmonic}
\delta\phi(t)\delta\Big[{\cal F}[\vec q,\vec p]\Big]\det(\frac{\delta {\cal F}}{\delta\phi})
\eeq
for some gauge fixing functional ${\cal F}:X\to \mathfrak{so}(2)$. The measure \eqref{Thom Harmonic} defines a conditional expectation from functions of $\{\vec{q}, \vec{p}, \phi \}$ to functions of $\{\vec{q}, \vec{p}\}$ as
\beq \label{Harmonic Cond Exp}
	\bigg(T(\mathfrak{F})\bigg)(\vec{\tilde q}, \vec{\tilde p}) = \int d\phi(t)\delta\Big[{\cal F}\circ a_\phi[\vec{\tilde q},\vec{\tilde p}]\Big]\det(\frac{\delta {\cal F}}{\delta \phi}) \mathfrak{F}(\vec{\tilde q}, \vec{\tilde p}, \phi).
\eeq
For example, we may take $\mathfrak{F}=i(f)$ with $f\in \Omega^0(X)$, in which case the conditional expectation \eqref{Harmonic Cond Exp} maps arbitrary functions on the phase space to gauge invariant observables as in \eqref{Cond exp of M_0}. 

To see how this works in this example, we will compose the path integral \eqref{Harmonic Path Integral} with the conditional expectation \eqref{Harmonic Cond Exp} to obtain a path integral on the full extended phase space. A nice choice for ${\cal F}$ is
\beq \label{Angle Fixing}
{\cal F}=\tan^{-1}(y(t)/x(t))
=\phi(t)+\tan^{-1}(\tilde y(t)/\tilde x(t))
\eeq
In this case the F-P determinant is unity. Writing for brevity
\beq
	\left[d\tilde{\gamma}(t)\right] = \left[d \tilde{p}_x(t)d \tilde{p}_y(t)d \tilde{x}(t)d \tilde{y}(t)\right],
\eeq
we obtain 
\beqn \label{Harmonic Extended Path Integral}
\tilde{\varphi}_t \circ i(f)
&=&Z_\varphi^{-1}\int \left[d \tilde{\gamma}(t) d\phi(t) dA_0(t) \right] e^{i\int dt\, L[\tilde q,\tilde p]}e^{-i\int dt\; A_0(t)\tilde L_z(t)}\delta[\phi+\tan^{-1}(\tilde y/\tilde x)] \Big(i(f)\Big)(\vec{\tilde q}, \vec{\tilde p},\phi)
\eeqn
To proceed, we change variables on phase space $(\tilde x,\tilde y,\tilde p_x,\tilde p_y)\to (\tilde r,\tilde p_r,\tilde \theta,\tilde p_{\theta})$. The Jacobian is trivial, and recognizing $\tilde L_z=\tilde p_\theta$, we obtain
\beqn
\tilde{\varphi}_t \circ i(f)
&=&Z_\varphi^{-1}\int \left[d \tilde{\gamma}(t) d\phi(t) dA_0(t)\right] e^{i\int dt\Big[\tilde p_ r\dot{\tilde r}-\frac12\tilde r^2-\frac{1}{2}\tilde p_ r^2+\tilde p_\theta\dot{\tilde\theta}-\frac{1}{2\tilde r^2}\tilde p_\theta^2-\tilde p_\theta A_0\Big]}\delta[\phi+\tilde\theta] f(\tilde r,\tilde p_r,\tilde p_{\theta},\tilde \theta+\phi)
\nonumber\\
&=&Z_\varphi^{-1}\int \left[d\tilde p_ r(t)d\tilde  r(t) d\phi(t) \right] e^{i\int dt\Big[\tilde p_ r\dot{\tilde r}-\frac12\tilde r^2-\frac{1}{2}\tilde p_ r^2\Big]}f(\tilde r,\tilde p_r,0,0)
\nonumber\\
&=&\tilde Z_\varphi^{-1}
\int \left[d\tilde p_ r(t)d\tilde  r(t)\right]e^{i\int dt\,\tilde p_ r\dot{\tilde r}}e^{-\frac12i\int dt(\tilde r^2+\tilde p_ r^2)} f([\vec q],[\vec p]).\label{radshoint}
\eeqn
We have also included in \eqref{Harmonic Extended Path Integral} an integral over the gauge field $A_0$ which serves to implement the constraint $\tilde p_{\theta} = 0$. The bracketed elements in \eqref{Harmonic Extended Path Integral} can be written explicitly, in this gauge, as
\beq
	[\vec{q}] = \begin{pmatrix}\tilde{r}  \cr 0\end{pmatrix}, \qquad [\vec p] = \begin{pmatrix}\tilde{p}_r \cr 0\end{pmatrix}.
\eeq
which are representatives of each gauge equivalence class as determined by the gauge fixing condition \eqref{Angle Fixing}. One should interpret $f([\vec q],[\vec p])$ as a dressed operator insertion. The path integral \eqref{Harmonic Extended Path Integral} reproduces the algebra of gauge invariant observables expected from the operator based analysis \eqref{Harmonic Gauge Invariant Algebra}. A small remaining subtlety here is that the coordinate $\tilde r$ is positive in \eqref{radshoint}; this can be treated by gauging spatial parity of an ordinary oscillator of frequency $\omega$. This gauging then removes the states of energy $E-E_0= n$ for odd $n$.

\subsection{Yang Mills}\label{sec: Yang-Mills}

As a second example, we consider Yang-Mills theory in 4d. In contrast to the quantum mechanics example of the last subsection, there is an important conceptual feature present here which is the existence of a gauge charge having support on surfaces of codimension two relative to spacetime. We will take care to keep track of this. We consider quantizing on a spatial hypersurface $\Sigma$; so the theory has a symplectic potential 
\beq \label{YM Sym potential}
\theta_\Sigma=\frac{1}{2g^2}tr\Big(\pi_j\delta A_j\Big)vol_\Sigma
\eeq
and a Hamiltonian\footnote{The ensuing analysis will reveal the importance of corner degrees of freedom. In this sense, the Hamiltonian presented here is not the most general possible as it could be supplemented with boundary terms.}
\beqn \label{YM Hamiltonian}
H=\frac{1}{2g^2} tr\Big(\pi_j\pi_j +B_{j}B_{j}\Big),\qquad B_i=\epsilon_{ijk}\pa_jA_k.
\eeqn
Both \eqref{YM Sym potential} and \eqref{YM Hamiltonian} are invariant under the hypersurface gauge transformation
\beq
A_i(\vec x)\mapsto V^{-1}(\vec x)(A_i(\vec x)+\pa_i)V(\vec x),\qquad \pi_i(\vec x)\mapsto V^{-1}(\vec x)\pi_i(\vec x)V(\vec x),
\eeq
where $V$ is regarded as an element of some compact group $G$. The symplectic form is non-degenerate off-shell, however going on-shell invokes the Gauss constraint leading to a degeneracy and necessitating a corresponding gauge fixing.

Although one could apply the techniques of RAQ or BRST to construct gauge invariant states and dressed operators, in this section we will discuss the path integral quantization more explicitly. To this effect we consider a $t$-parameterized curve through phase space and begin by promoting the hypersurface gauge invariance to be local along such curves. In particular, this means that the phase space action (or Hamilton's principle function) takes the form
\beq \label{YM Action 1}
S=\frac{1}{g^2}\int dt \int vol_\Sigma\; tr\Big( \vec\pi\cdot\dot{\vec A}-\Big[\frac12\vec\pi^2+\frac12 \vec B^2+\vec \pi\cdot\vec DA_0\Big]\Big) 
\eeq
As in the previous example, $A_0$ covariantizes the action with respect to $t$-dependent gauge transformations, here in a way that is consistent with the additional hypersurface gauge invariance. That is, the field $A_0$ has a hypersurface covariant derivative
\beq
D_jA_0=\pa_j A_0+\big[A_j,A_0\big].
\eeq
We note in passing that the resulting theory \eqref{YM Action 1} possesses a global Lorentz symmetry. 

Expanding \eqref{YM Action 1} we obtain the action
\beq  \label{YM Action 2}
S=\frac{1}{g^2}\int dt \int vol_\Sigma\; tr\Big( \vec\pi\cdot\dot{\vec A}-\Big[\frac12\vec\pi^2+\frac12 \vec B^2+\pa_j(A_0\pi_j )-A_0(\pa_j\pi_j +[A_j,\pi_j])\Big]\Big) 
\eeq
Up to a total derivative, $A_0$ plays the role of a Lagrange multiplier\footnote{Often one encounters a discussion of $A_0$ as being canonically conjugate to a momentum field $\pi_0$, with a first class constraint $\pi_0=0$. In this exposition, it serves no purpose to do so.} for the Gauss constraint $G=D_j\pi_j=\pa_j\pi_j +[A_j,\pi_j]$. More precisely, $A_0$ is a Lagrange multiplier in the bulk of the hypersurface. In standard analyses this is often taken to be the end of the story, but we are interested in evaluating a path integral for a generic subregion of the hypersurface $\Sigma$. To do so, we must introduce an embedding map $\phi_{(2)}:S\to\Sigma$ for a corner, regarded here as the boundary of the subregion.\footnote{Note in this theory, we are not gauging diffeomorphisms and so we do not expect generally that the embedding map will become dynamical. In fact, as explained in Section 2, we expect that a Lie algebroid isomorphism becomes physical at a boundary. Here, we are not using that language, but we will in fact arrive at that interpretation shortly. In particular we will find that gauge transformations at the boundary become physical global symmetries, inducing a non-trivial boundary symplectic form which couples the Noether gauge charge to the boundary Maurer-Cartan form (which in the algebroid language is the vertical part of the connection). That is, the Lie algebroid morphism is determined by the embedding $\phi_{(2)}$ and $u$. } Then \eqref{YM Action 2} becomes
\beq
S=\frac{1}{g^2}\int dt \Big(\int_\Sigma vol_\Sigma\; tr\Big( \vec\pi\cdot\dot{\vec A}-\Big[\frac12\vec\pi^2+\frac12 \vec B^2-A_0G)\Big]\Big) -\int_{S}vol_S\,tr(a_0\chgdens{})\Big)
\eeq
where $a_0$ is $A_0\circ\phi_{(2)}$ and $\chgdens{}=\phi_{(2)}^*(*_3\pi)$, the pullback of $\pi_j$ to the corner. As we shall see, the inclusion of these corner degrees of freedom has important implications for the physics of the subregion.

The naive unextended phase space of the Yang-Mills theory is coordinatized by $\{A_i,\pi_i\}$. In the absence of a boundary, we would define an extended phase space as a principal bundle over the naive phase space and we would manage the corresponding theory through the usual Faddeev-Popov procedure. On the other hand, in the presence of a boundary, physical fields will emerge with support on the boundary that survive the gauge-fixing procedure. This corresponds to the existence of a residual global symmetry at codimension-$2$, with gauge-fixed fields continuing to transform in non-trivial representations of the resulting global symmetry group. In Section 2, we described this in general through the idea of separate codimension-1 and codimension-2 extensions to the phase space. 
We will describe how this idea plays out in the specific case of the Yang-Mills theory.

To begin we consider the group action on the space of fields,
\beq\label{YMintroU}
A_j=U^{-1}\tilde A_jU+U^{-1}\pa_jU,\qquad \pi_j=U^{-1}\tilde \pi_j U,\qquad A_0=U^{-1}\tilde A_0U+U^{-1}\dot U.
\eeq
The principle function satisfies $S[A_i,\pi_i,A_0]=S[\tilde A_i,\tilde\pi_i,\tilde A_0]\equiv \tilde S$, independent of $U$. As in the previous example, we can interpret the tilded variables as defining a gauge orbit, and in that sense, the gauge symmetry can be thought of as a right action on $U$ alone with
\beq
U\mapsto UV,\qquad \tilde A_j\mapsto \tilde A_j,\qquad \tilde \pi_j\mapsto \tilde\pi_j.
\eeq
Consequently we refer to $(\tilde A_j,\tilde\pi_j)$ as dressed fields. One can check that other quantities, such as the magnetic field, are similarly dressed. 

%

Using \eqref{YMintroU} the symplectic potential density becomes
\beqn
\theta_\Sigma
&=&\frac{1}{2g^2}tr\Big(\tilde\pi_j(\delta\tilde A_j +[\tilde A_j,\delta UU^{-1}]-\delta UU^{-1}\pa_jUU^{-1}+\pa_j\delta UU^{-1}) \Big)vol_\Sigma\nonumber\\
&=&\frac{1}{2g^2}tr\Big(\tilde \pi_j\delta\tilde A_j+\pa_j\Big[\delta U U^{-1}\tilde \pi_j\Big]
  - \delta UU^{-1}\tilde G\Big)vol_\Sigma\label{splitthetaym}
\eeqn
where $\tilde G=\pa_j\tilde \pi_j+[\tilde A_j,\tilde\pi_j]$ is the dressed Gauss law constraint. 
Integrating over $\Sigma$, we then find the symplectic potential
\beq \label{YM Symplectic Potential}
\Theta_\Sigma=\frac{1}{2g^2}\int_\Sigma vol_\Sigma\, tr\Big(
\tilde \pi_j\delta\tilde A_j
-\delta UU^{-1}\tilde G\Big)
  +\frac{1}{2g^2}\int_S vol_S\, tr( \delta u u^{-1}\tilde{\chgdens{}})
\eeq
In \eqref{YM Symplectic Potential} we can recognize the Maurer-Cartan form $\varpi_{(1)}=\delta U U^{-1}$. In fact, \eqref{YM Symplectic Potential} has the same form as the general expression \eqref{charge piece} with  the constraint $\constr{}=-*_3\tilde G$. Moreover, in the presence of a boundary the charge $\tilde{\chgdens{}}$ appears in the symplectic potential conjugate to the boundary value of the Maurer-Cartan form $\varpi_{(2)}=\delta u u^{-1}$. We interpret this to mean that there is a corner symplectic pair present. The charge $\tilde{\chgdens{}}$ acts as the generator of the emergent global symmetry supported on the boundary, and the $u$ fields are subsequently in a representation of this global symmetry.

Eqn. \eqref{YM Symplectic Potential} implies that we should regard the physical phase space to include the fields\footnote{Here we have separated the boundary value $u$ from the bulk transformation $U$. In this sense, the bulk gauge transformation $U$ should be regarded as going to the identity at the boundary of the subregion.} $X=\{A_i,\pi_i,u,\chgdens{}\}$. In the language of Section \ref{sec: QEPS} this constitutes what we referred to as the extension at codimension-$2$. Hereafter we introduce the notation $\gamma = (A_i, \pi_i, u, \chgdens{})$ to refer collectively to these degrees of freedom. The gauge group acts on $X$ via the map $a^{(1)}: G \times X \rightarrow X$ defined by \eqref{YMintroU} in the bulk along with a trivial action on $u,\chgdens{}$. We have included a superscript $(1)$ to distinguish the gauge symmetry $a^{(1)}$ from the global symmetry generated by the corner charge. In the following we will refer to the latter by $a^{(2)}$.  
For an arbitrary function $f(\gamma)$ on $X$, we can define the map \eqref{inclusion of M_0 in M_ext} which passes $f$ to a function on the extended phase space
\beq
	\bigg(i(f)\bigg)(\tilde{\gamma}, U) = f \circ a^{(1)}_{U}(\tilde{\gamma})=f(\gamma). 
\eeq 

The codimension-1 piece of \eqref{splitthetaym}, $tr(\tilde G\varpi )$, resembles a pure gauge symplectic pair. 
In order for the above analysis to correspond to a canonical transformation for fields in the bulk of the hypersurface, we introduce a gauge fixing to project $tr(\tilde \pi_j\delta\tilde A_j)$ to $X/G$. In the path integral this corresponds to the introduction of a conditional probability distribution
\beq \label{Thom Harmonic YM}
\Phi(U|\tilde{\gamma})=\delta\big({\cal F}\circ a^{(1)}_U(\tilde{\gamma})\big)\Delta_{\cal F}\circ a^{(1)}_U(\tilde{\gamma})
\eeq
for some ${\cal F}:X\to \mathfrak{g}$.    
 The measure \eqref{Thom Harmonic YM} defines a conditional expectation from functions of $\{A_i,\pi_i,u,\chgdens{},U \}$ to functions of $\{A_i,\pi_i,u,\chgdens{}\}$ as
\beq \label{YM Cond Exp}
	\big(T(\mathfrak{F})\big)(\tilde{\gamma}) = \int \mu(U(t))\delta\big({\cal F}\circ a^{(1)}_U(\tilde{\gamma})\big)\Delta_{\cal F}\circ a^{(1)}_U(\tilde{\gamma}) \mathfrak{F}(\tilde{\gamma}, U).
\eeq
with $\mu(U(t))$ a left-invariant Haar measure on $G$ in the bulk. A representative gauge choice is ${\cal F}=D_j A_j$. At least in the Abelian case, the $A_i,\pi_i$ are then  reduced to their transverse components.

We now have all the pieces we need to construct a path integral for the Yang-Mills theory. In particular\footnote{We are not being specific about the details of the expectation value computed in \eqref{Path integral for YM}, which would be determined by specifying precisely the boundary conditions on the measure, as well as the details of the insertion.},
\beq \label{Path integral for YM}
	\tilde{\varphi}_t \circ i(f) \equiv Z^{-1}_{\tilde{\varphi}} \int [d\tilde{\gamma}(t) \mu(U(t)) dA_0(t)] e^{i \tilde{S}_0} e^{-i \frac{1}{g^2} \int dt \wedge \int_{\Sigma}vol_{\Sigma}\, A_0 \tilde{G}} \Phi(U \mid \tilde{\gamma}) \bigg(i(f)\bigg)(\tilde{\gamma}, U)
\eeq
The action $\tilde{S}_0$ is defined through the equation
\beq \label{YM action}
\tilde	S =\tilde S_0 - \frac{1}{g^2} \int dt \int_{\Sigma}vol_\Sigma\, tr( A_0 \tilde G),
\eeq
that is we have separated the bulk Lagrange multiplier away from the physical part of the action. The integral \eqref{Path integral for YM} can be evaluated in a series of steps. First, integrating over $A_0$ generates a delta functional which sets the constraint current $\tilde{G} = 0$. Second, integrating over $U$ activates the gauge fixing. In the end, we are left with an integral of the form
\beq
	\tilde{\varphi}_t \circ i(f) \equiv \tilde Z^{-1}_{\varphi} \int \bigg[d[\gamma(t)]\bigg] e^{i\tilde{S}_0} f([\gamma]),
\eeq
where $[\gamma]$ is the representative of the gauge orbit which is selected by the chosen gauge fixing. 

An important observation that should be made is that the gauge fixed function $f([\gamma])$ may still depend on data derived from the corner fields, $(u, \chgdens{})$. In other words, gauge fixed functions remain in the representation space of the global symmetry $a^{(2)}$. Thus, while implementing the constraints of the theory removes the redundancy introduced by gauge symmetry, the algebra of gauge invariant operators retains a crossed product structure. Here, the `extension' is provided by corner supported transformations which act as global symmetries in the gauge fixed theory. Formally, one may therefore interpret the path integral \eqref{Path integral for YM} as including a sum over corner supported Lie algebroid isomorphisms parameterized by the field $u$. In the gravitational context the analogous isomorphism will correspond to the corner embedding \cite{Ciambelli:2021nmv,Ciambelli:2021vnn,Ciambelli:2022cfr} as was implemented in a quantum context in recent work \cite{Balasubramanian:2023dpj}. We will have more to say about this in the next section.

\subsection{$4d$ Gravity on a Null Hypersurface} \label{sec: Gravity}

In this penultimate section, we consider the phase space of a gravitational theory (in particular, the Einstein-Hilbert theory) formulated on a null hypersurface ${\cal N}$. This will serve as an example of a diffeomorphism-invariant theory. The intrinsic geometry of such a null hypersurface is given by a Carroll structure and a Carrollian connection.  A Carroll structure is determined by a codimension-1 metric $q_{ab}$ and a null vector field $\ell^a$. An Ehresmann connection $k_a$ yields a rigging, defining a horizontal section of the tangent bundle and a corresponding projector $q_a{}^b= \delta_a{}^b - k_a\ell^b$, $\ell^a$ being vertical. 
As reviewed recently in \cite{Ciambelli:2023mir}, the symplectic potential \cite{Chandrasekaran:2021hxc} (see also \cite{Freidel:2022bai,Freidel:2022vjq}) can be written as
\beq\label{thcan}
\Theta^{\mathsf{can}}=\int_{\cal N} \varepsilon_{\cal N}\Big(\frac{1}{2}
\tau^{ab}\delta q_{ab}
-\tau_a\delta \ell^a\Big).
\eeq
where 
\beqn
\kappa_N^2\,\tau_a=  \pi_a-\theta k_a,\qquad
\kappa_N^2\, \tau_a{}^{b}=\sigma_a{}^{b}-\mu q_a{}^{b}.
\eeqn
Here we are using the same notation as \cite{Ciambelli:2023mir}, and all of the details and associated discussion can be found there. Our goal in this section is to draw out an interpretation of the analysis of \cite{Ciambelli:2023mir} that coincides with the general picture presented in this paper. As in \cite{Ciambelli:2023mir}, we will focus on two of the gauge symmetries, namely an internal boost symmetry and the diffeomorphisms that reparameterize the null coordinate, leaving the general case to future work. 

To begin, we can transform the fields of the theory by a boost parameterized by a function $\alpha$ such that $
\delta \ell^a=-\delta\alpha \ell^a-e^{-\alpha}\delta u^a$ with $k_a\delta u^a=0$. Under this transformation, one finds that the canonical symplectic potential can be written
\beq\label{nullgravgensympot}
\kappa_N^2\Theta^{\mathsf{can}}=\int_{\cal N} \varepsilon^{(0)}_{\cal N}(\frac12 \Omega_{(\alpha)}\sigma_{(\alpha)}^{ab}\delta  \overline{q}^{(\alpha)}_{ab}-\mu_{(\alpha)}\delta  \Omega_{(\alpha)}+\Omega_{(\alpha)}\pi^{(\alpha)}_a\delta u_{(\alpha)}^a)
-\int_{\cal C} \varepsilon^{(0)}_{\cal C}\bar\Omega_{(\alpha)}\delta\alpha
\eeq
Here we have separated out the determinant of the metric, $q_{ab}=\Omega\bar q_{ab}$ with $\det\bar q=1$, and $\Omega=\sqrt{\det q}$. The final term in \eqref{nullgravgensympot} is a corner contribution with $\bar\Omega=\Omega\circ \phi_{(2)}$, where $\phi_{(2)}$ is an embedding of the corner ${\cal C}$ in the hypersurface. The sub/superscripts $(\alpha)$ in \eqref{nullgravgensympot} are meant to remind the reader that the fields appearing therein have been transformed by a local boost transformation.\footnote{Since we consider multiple gauge symmetries in this section, a more precise notation is needed here, replacing the tilde notation of previous sections. We henceforth drop the sub/superscripts.} 
 In fact, the result \eqref{nullgravgensympot} can be thought of in terms of extracting the internal boost parameter as a canonical variable, with the rest of the fields being interpreted as defining a point on the boost orbit, the analogue of \eqref{rotateSHO} in the harmonic oscillator or \eqref{YMintroU} in Yang-Mills. Given that interpretation, we expect to see the boost constraint appear in the bulk of the hypersurface and the boost charge on the corner. In this specific case, it happens that the boost constraint vanishes identically off-shell, and so the boost current appears entirely as the boost charge on the corner. One sees from \eqref{nullgravgensympot} that the boost corner charge is given  by the area density $\bar\Omega$. So $(\alpha,\bar\Omega)$ can be considered as a non-trivial symplectic pair with support on a corner, that is, an edge mode pair.

Next, we consider a reparameterization of the null fibre. We will do so in such a way that extracting the corresponding gauge transformation does not impact the previous boost symmetry analysis. To this end, we consider a combination of a diffeomorphism and a boost; this can be thought of as `block-diagonalizing' the gauge transformations, and in \cite{Ciambelli:2023mir} was referred to as a primed diffeomorphism. The primed diffeomorphism is quantified by a standard diffeomorphism $v\mapsto V(v,\sigma)$ (referred to as the dressing time) together with a compensating boost. It is convenient at this point to introduce $U:=\pa_vV$, which can be thought of as the boost group element associated with the primed diffeomorphism. In particular for an infinitesimal diffeomorphism $V(v,\sigma) \simeq v + \phi(v,\sigma)$ where $\phi$ is to be regarded as the vector field generating the diffeo, we have $U\simeq 1+\pa_v\phi(v,\sigma)$, see Eqn. (148) in \cite{Ciambelli:2023mir} for more detail. 

As was the case in the previous examples, our intention will be to interpret the action of the primed diffeomorphisms as a canonical transformation on the space of fields. To this end, we introduce the bulk action $a_V$ which transforms the fields by a combination of the pullback by $V$ and a boost $U$. In particular,
\beq \label{Primed diffeo action}
	a_V(\tilde\Omega,\tilde\mu,\tilde{\bar q}_{ab}, \tilde\sigma_a{}^b)=(\tilde\Omega\circ V,\; U(\tilde\mu\circ V)+U^{-1}\pa_v U,\;\tilde{\bar{q}}_{ab}\circ V,\; U\tilde\sigma_{ab}\circ V).
\eeq
By extension, we have for a phase space function $F$ the dressing
\beq \label{Primed diffeo dressing}
i(F)(\tilde\Omega,\tilde\mu,\tilde{\bar q}_{ab}, \tilde\sigma_a{}^b,V)=F\circ a_V(\tilde\Omega,\tilde\mu,\tilde{\bar q}_{ab}, \tilde\sigma_a{}^b)=F(\tilde\Omega\circ V,\; U(\tilde\mu\circ V)+U^{-1}\pa_v U,\;\tilde{\bar{q}}_{ab}\circ V,\; U\tilde\sigma_{ab}\circ V)
\eeq
In Eqn. \eqref{Primed diffeo action} and \eqref{Primed diffeo dressing} we have introduced tilded variables with an interpretation analogous to that which was described below Eqn. \eqref{rotateSHO} and \eqref{YMintroU}. 

Applying the transformation \eqref{Primed diffeo action} to the symplectic potential (simplifying notation somewhat), we find
\beqn\label{nullgravgensympottilde}
\kappa_N^2\Theta^{\mathsf{can}}&=&\int_{\cal N} \varepsilon^{(0)}_{\cal N}(\frac12 \tilde\Omega U\tilde\sigma^{ab}(\delta\tilde q_{ab}+\pa_V\tilde q_{ab}\delta V)-\Big(U \tilde{\mu} + U^{-1}\pa_vU\Big)(\delta\tilde\Omega+\pa_V\tilde\Omega\delta V)+\tilde\Omega\tilde\pi_a\delta u^a)
-\int_{\cal C} \varepsilon^{(0)}_{\cal C}\Omega\delta\alpha
\nonumber\\
&=&\int_{\cal N} \varepsilon^{(0)}_{\cal N}\Big( U(\kappa_N^2\tilde\theta^{\mathsf{can}}+\tilde{C}\delta V)
-U\pa_V^2\tilde\Omega\delta V
-U^{-1}\pa_vU\delta\Omega
\Big)
-\int_{\cal C} \varepsilon^{(0)}_{\cal C}\Omega\delta\alpha
\nonumber\\
&=&\int_{\cal N} \varepsilon^{(0)}_{\cal N}\Big( U(\kappa_N^2\tilde\theta^{\mathsf{can}}+\tilde{C}\delta V)
-\delta\Big[\Omega U^{-1}\pa_vU\Big]
-U\pa_V^2\tilde\Omega\delta V
+\pa_v\Big[ U^{-1}\delta U\Big]\Omega
\Big)
-\int_{\cal C} \varepsilon^{(0)}_{\cal C}\Omega\delta\alpha
\nonumber\\
&=&\int_{\cal N} \varepsilon^{(0)}_{\cal N}\Big( U(\kappa_N^2\tilde\theta^{\mathsf{can}}+\tilde{C}\delta V)
-\delta\Big[\Omega U^{-1}\pa_vU\Big]
+\pa_v\Big[ \Omega U^{-1}\delta U-\delta V\pa_V\tilde\Omega\Big]\Big)
-\int_{\cal C} \varepsilon^{(0)}_{\cal C}\Omega\delta\alpha
\nonumber\\
&=&\int_{\cal N} \tilde\varepsilon^{(0)}_{\cal N} (\kappa_N^2\tilde\theta^{\mathsf{can}}+\tilde{C}\delta V\circ V^{-1})\circ V
+\int_{\cal C} \varepsilon^{(0)}_{\cal C}\Big[ \tilde\Omega U^{-1}\delta U-\pa_V\tilde\Omega\delta V\circ V^{-1}
-\tilde\Omega\delta\alpha\Big]\circ V
\eeqn
where in the last line we dropped a total variation and we have defined  
\beq\label{rmu}
\tilde{C}=\pa_V^2\tilde\Omega-\tilde\mu\pa_V\tilde\Omega+ \tilde\Omega\left( \tilde\sigma^2+\kappa_N^2\,T_{VV}^{mat}\right).
\eeq
We recognize this as the Raychaudhuri constraint, written in the tilde variables.
As established in \cite{Ciambelli:2023mir},  the Raychaudhuri constraint $\tilde C$ is conjugate to $V$ -- it is the constraint appearing in the bulk term of the primed diffeomorphism Noether current. We moreover see that the transformation \eqref{Primed diffeo action} has brought to light two additional terms in the corner piece of the symplectic potential. We now interpret the first two terms in the corner term of \eqref{nullgravgensympottilde} as corresponding to the charge of the primed diffeomorphism, while the third term is related to the boost charge. Notice that these terms are of the expected form $tr(\chgdens{}\varpi^{(2)})$, for appropriate Maurer-Cartan forms $\varpi^{(2)}$, $U^{-1}\delta U$ for the boost part of the primed diffeomorphism, $\delta V\circ V^{-1}$ for the standard diffeomorphism, and $\delta \alpha$ for the internal boost.

To interpret \eqref{Primed diffeo action} as a canonical transformation a gauge-fixing condition is required. The transformed canonical symplectic potential $\tilde\theta^{can}$ contains the spin-$0$ term $-\tilde\mu \delta\tilde\Omega$. In  \cite{Ciambelli:2023mir}, the (classical) condition $\tilde\mu=0$ was chosen; here we can promote this to a Fadeev-Popov gauge fixing procedure in a quantum theory by choosing an appropriate gauge-fixing functional ${\cal F}$ (both for boosts and primed diffeomorphisms). 

Since we are considering the theory from the extended phase space point of view, the algebra of constraints and the algebra of charges are expected to close. As explained in \cite{Ciambelli:2021nmv,Ciambelli:2021vnn}, this comes about via the proper inclusion of embedding maps (or Lie algebroid morphisms in the more general presentation of \cite{Klinger:2023qna}) in the phase space. In \cite{Ciambelli:2023mir}, this was shown explicitly for the constraint algebra, where it became clear that the spin-$0$ sector can be interpreted as codimension-1 extended degrees of freedom. This will also be true for the spin-1 sector, associated with the Damour constraint. For the charge algebra, however, a codimension-2 extension is required, and the above analysis can be interpreted in those terms. In fact the Maurer-Cartan forms seen above can be recognized formally as the vertical part of a connection on the configuration algebroid associated with boost and diffeomorphism gauge symmetries.\footnote{More precisely, we are addressing only a portion of this algebroid, as we are not considering general diffeomorphisms or the full set of internal symmetries.} The diffeomorphism $V$ can in fact be thought of as an embedding of the corner inside the null hypersurface -- it defines a cut or section of the Carroll structure. While $V$ is defined in the bulk of the hypersurface, it is pure gauge in the sense that it is conjugate to a constraint; there is a residual part of $V$ that has support on a corner. This generates a subgroup of the universal corner symmetry.

Because the extended phase space construction yields integrability of charges, we expect that contracting the symplectic potential with the phase space vector field generating symmetries yields the corresponding charge. Working on-shell, in the case of the boost this immediately yields
\beq
G_\lambda = \frac{1}{\kappa_N^2}\int_{\cal C}\varepsilon^{(0)}_{\cal C}\tilde\Omega \lambda
\eeq
where we used $I'_{\hat\lambda}\delta\alpha=-\lambda$, while for the primed diffeomorphism
\beq
Q'_f= \frac{1}{\kappa_N^2}\int_{\cal C}\varepsilon^{(0)}_{\cal C}\left(\tilde\Omega \pa_v f-f\pa_v \tilde\Omega\right)
\eeq
where we used $I'_{\hat f}\delta V=f$ and $I'_{\hat f}U^{-1}\delta U=\pa_vf$. These expressions coincide with the charges found in \cite{Ciambelli:2023mir} by the more standard approach of constructing Hamiltonian functions relative to the symplectic form. 

The analysis of this section should be regarded as providing the skeleton for the quantization of gravity on null hypersurfaces along with an actionable prescription for implementing the requisite diffeomorphism constraints (although here and in \cite{Ciambelli:2023mir} only the Raychaudhuri constraint has been considered). We emphasize that this has been done without recourse to an observer, with the spin-0 sector of the naive phase space playing the role of a clock degree of freedom relative to which generic observables can be gravitationally dressed. As an additional point of emphasis, we should note that the appearance of the area as a Noether charge for corner supported boosts seems to signal a connection with the generalized entropy. We plan to address this in detail in future work.

We have stopped short of writing down the path integral in this section in recognition of the fact that there are important technical challenges which warrant a more complete discussion to be presented in future work. For example, integrating over the group of gauge transformations in this gravitational context raises the sticky issue of the non-locally compact nature of the group of diffeomorphisms. Nevertheless, as presented in Appendix E of \cite{Klinger:2023tgi}, it seems possible that this difficulty may be circumvented by treating the set of diffeomorphisms as the pair groupoid and using the fact that every Lie groupoid is locally compact as a topological groupoid. Then one can appeal to the theory of left invariant Haar systems for locally compact groupoids which provides a suitable generalization to the theory of left invariant Haar measures in the context of a locally compact group. This observation should also prove instrumental in making sense of integrating over the set of all corner embeddings, as implicated by the codimension-2 aspect of the null hypersurface story discussed in this section. This is the analog of integrating over the field $u$ in the Yang-Mills example.

\section{Discussion} \label{sec: discussion}

In this note we have made use of the crossed product to study the quantization of generic gauge theories, as well as the closely associated problem of implementing constraints. A crucial ingredient in this undertaking has been the extended phase space, which figures into the quantization of gauge theories through the correspondence \eqref{commutative diagram}. Using this correspondence and the geometry of Atiyah Lie algebroids we are able to formulate an integrable symplectic structure in which charges associated with internal gauge symmetries and diffeomorphisms are treated on  the same footing. 

In Section \ref{sec: preliminaries} we provided a sketch of the operator algebra which arises if one can adequately quantize the aforementioned symplectic geometry. An important insight that is drawn from this analysis is that this algebra can be regarded as the result of a pair of crossed products associated with automorphisms generated by charges that live on codimension one and codimension two submanifolds of spacetime, respectively. The codimension one charges generate gauge transformations and therefore must be implemented as constraints in the physical theory. By contrast, the codimension two charges are unconstrained and therefore generate physical transformations of states in the quantum theory. These charges are a central ingredient in  the corner proposal \cite{Ciambelli:2021vnn,Ciambelli:2022vot,Ciambelli:2022cfr,Donnelly:2016auv,Freidel:2020xyx,Freidel:2020svx,Freidel:2020ayo,Freidel:2021cjp,Freidel:2023bnj,Speranza:2017gxd,Donnelly:2022kfs,Speranza:2022lxr,donnelly2021gravitational}, and play a key role in regulating entanglement observables in diffeomorphism covariant theories \cite{Klinger:2023tgi}. 

In Section \ref{sec: con-quant} we have undertaken a general study of implementing constraints for a quantum theory in which a von Neumann algebra $\mathcal{M}$ is acted upon by an automorphism $\alpha: G \times \mathcal{M} \rightarrow \mathcal{M}$ regarded as a gauge symmetry. Our main goal in this section was to emphasize that the crossed product $\mathcal{M} \rtimes_{\alpha} G$ is the correct algebraic setting in which to accomplish this. In Subsections \ref{sec: RAQ}, \ref{sec: BRST}, \ref{sec: Path}, and \ref{sec: comm thm} we have provided an overview of four independent methods for implementing constraints in a quantum theory. In each subsection, we have highlighted precisely how the crossed product and the extended phase space figure naturally into the specific methodology at hand. For convenience, we review these observations now.

In the Hilbert space approach to Refined Algebraic Quantization, the role of the crossed product is to construct an auxiliary Hilbert space in which the constraints are automatically realized as unitary operators \eqref{ext0 rep}. From this point it is straightforward to follow the standard approach to obtain a physical Hilbert space in which all states are invariant under the gauge group. Alternatively, from the operator algebraic perspective the crossed product admits a projection to a subalgebra of $G$-invariant operators \eqref{Group averaging in algebra}. The GNS Hilbert space associated with this subalgebra can play the role of a physical Hilbert space. The relationship between these approaches is summarized by the commutative diagram:
\begin{equation} \label{RAQ commutative diagram}
\begin{tikzcd}
\mathcal{M} \rtimes_{\alpha} G \arrow[r, "GNS"] \arrow[d, "\Pi_d"]
& \mathcal{H}_{aux} \arrow[d, "\eta"] \\
\mathcal{M}_{inv} \arrow[r, "GNS"]
& \mathcal{H}_{phys}
\end{tikzcd}
\end{equation} 
where here $\Pi_d$ is the group averaging map in the algebra, and $\eta$ is the rigging map. Of central importance in the algebraic approach to RAQ is the existence of a tracial weight in the crossed product algebra which plays the role of an invariant vacuum. This emphasizes the interplay between the two roles of the crossed product that we have stressed, in order to allow for the implementation of constraints it is necessary for the algebra to first be rendered semi-finite.

In the BRST quantization scheme, the role of the crossed product is to furnish a representation of the BRST algebra complete with ghost and antighost degrees of freedom \eqref{Ghosts and Antighosts}. This is made possible by constructing a covariant representation of the crossed product on the tensor product space $\Omega^{\bullet}(G)\otimes \mathcal{H}$ where $\mathcal{H}$ is a standard representation of the algebra and $\Omega^{\bullet}(G)$ is the exterior algebra of the group $G$. A Lie algebra representation on this Hilbert space is a tensor product between a standard representation on $\mathcal{H}$ and the co-adjoint representation acting on the exterior algebra generators \eqref{Extended Lie algebra Rep}. The BRST differential is obtained naturally by acting on elements of the Hilbert space with the extended representation of the ghost field $c = c^A \otimes \un{t}_A$ \eqref{Extended Lie algebra Rep}. Using the Cartan algebra inherited from the exterior algebra \eqref{BRST relation}, it is easy to show that elements of the BRST cohomology are invariant under the extended action of the group and thus distinct BRST cohomology classes may play the role of a physical Hilbert space \eqref{BRST Invariance}.  

From the phase space perspective, a path integral is a weight on the Poisson algebra of a symplectic manifold \eqref{Sketch of path integral}. The problem of implementing constraints is motivated by the observation that generic observables (elements of the Poisson algebra) are not invariant under the action of the group $G$ on $X$. To remedy this, we pass to the extended phase space, $X_{ext} \simeq X \times G$, and introduce a conditional expectation which projects from the Poisson algebra of $X_{ext}$ to the Poisson algebra of $X$ \eqref{cond exp}. When this conditional expectation is paired with a dressing procedure \eqref{inclusion of M_0 in M_ext} the result is a map from the Poisson algebra of $X$ into the $G$-invariant sector therein \eqref{Cond exp of M_0}. Using this data the naive path integral can be promoted to a $G$-invariant path integral by including a gauge fixing delta function analogous to the standard approach of Faddev-Popov. Utilizing the correspondence \eqref{commutative diagram}, the Faddeev-Popov gauge fixing procedure can be implemented in a completely algebraic fashion. This correspondence is made particularly transparent by appealing to Haagerup's construction of a $C^*$ algebra whose weak closure yields the crossed product \cite{Haagerup1978I}. From this perspective, the gauge fixed path integral is analogous to a dual weight composed with a dressing procedure appropriate to Haagerup's algebra \eqref{Embedding}. 

Finally, we have the commutation theorem. From an algebraic point of view the commutation theorem may be regarded as an alternative definition of the crossed product. It states that, given a covariant system $(\mathcal{M},G,\alpha)$, the associated crossed product, $\mathcal{M} \rtimes_{\alpha} G$, is  precisely the invariant subalgebra of $\mathcal{M} \otimes B(L^2(G;\mu)$ under the extended automorphism $\theta \equiv \alpha \circ \text{Ad}_r$. Here $r: G \rightarrow U(L^2(G;\mu))$ is the right regular representation of $G$ on $L^2(G;\mu)$. From the point of view of the correspondence \eqref{commutative diagram}, the extended action $\theta$ is parallel to the complete right action of the group $G$ on the extended phase space \eqref{Full Right action EPS} which combines the $G$-action on the non-extended phase space, $a$, with the standard right action of $G$ on itself. The commutation theorem follows essentially from the fact that operators in the crossed product are `dressed'. When acting on dressed operators the original automorphism $\alpha$ can be absorbed entirely into a right translation of the dressing group elements. This translation is subsequently canceled by the explicit right action of the group on the dressed operator. In this way the two actions on the crossed product algebra conspire to leave all of its elements invariant \eqref{Invariance via commutation theorem}. The explicit right action of the group may be interpreted in analogy with the gauge fixing projection appearing in the path integral approach \eqref{Dressing}. Indeed, the role of this projection is precisely to offset any gauge variation with a compensating translation that solders each operator to a chosen reference point along its gauge orbit. 

The algebraic perspective on constraint quantization makes it  clear that each of these four approaches are merely devices for constructing conditional expectation-like objects which map operators in the crossed product to a $G$-invariant subalgebra. The need to relax some of the properties of conditional expectations as they are strictly defined in this context is useful because it provides some indication of how one may overcome pitfalls that have plagued constraint quantization, particularly in the gravitational context. For example, the group averaging construction yields a conditional expectation only when the group in question is compact, else it diverges and results in an operator-valued weight. However, as was discussed in Section \ref{sec: RAQ}, the group averaging map is just one possible choice for the requisite projection. A more flexible construction known as the generalized conditional expectation presents a possible alternative which always exists given a von Neumann algebra $\mathcal{M}$ with subalgebra $\mathcal{M}_0$ and faithful semi-finite normal weight $\varphi$ on $\mathcal{M}$. Every von Neumann algebra possesses a faithful semi-finite normal weight \cite{takesaki2003theory}, thus the only remaining question is whether the $G$-invariant subspace of $\mathcal{M}$ is a bona-fide von Neumann subalgebra. Nevertheless, while this would establish the existence of such a generalized conditional expectation it leaves the formidable problem of constructing it. What's more, as we have discussed in Section \ref{sec: con-quant} generalized conditional expectations generically fail to possess the bi-module property of an operator-valued weight. The repercussions of this failure deserves careful consideration relative to the problem of constructing and interpreting physical states/algebras. We plan to address this point in future work. 

A second problem that is encountered in constraint quantization for gauge theories and gravity is the fact that the constraints are not formulated in terms of a global group, but rather are local in spacetime. This problem is also addressed by our construction in passing the geometric structure of the constraints into the form of a groupoid. Again, however, there are technical subtleties which must be addressed in relation to, for example, integrating over Lie groupoids.\footnote{In forthcoming work, we address the problem of upgrading the crossed product to accommodate groupoid automorphisms. This utilizes the concept of a left Haar system belonging to a groupoid which generalizes the canonically defined Lebesgue integration on a group. See also related comments in \cite{Klinger:2023tgi}.}
We intend to explore these issues in forthcoming publications.

\appendix
\renewcommand{\theequation}{\thesection.\arabic{equation}}
\setcounter{equation}{0}

\section*{Acknowledgments}

We thank Luca Ciambelli, Laurent Freidel, Samuel Goldman, Weizhen Jia, and Pin-Chun Pai for useful conversations and collaborations on related projects. RGL acknowledges the support of Perimeter Institute for Theoretical Physics where part of this research was carried out. This work was partially supported by the U.S. Department of Energy under contract DE-SC0015655.

\section{Review of Atiyah Lie algebroids for Gauge Theories} \label{app: LA review}

As is usual in studying a gauge theory, our starting point is a principal $G$-bundle over a $d$-dimensional manifold $M$, $P(M,G)$. Here $G = \gravgroup \times \gaugegroup$, where $\gravgroup \subset Gl(d,\mathbb{R})$ and $\gaugegroup$ is some internal gauge group such as $SO(n)$. In other words, we are considering a theory with dynamical gauge fields and gravity. Associated with the principal bundle $P$ we have the Atiyah Lie algebroid $A = TP/G$. Recall that a Lie algebroid is a vector bundle over $M$ with a bracket $[\cdot,\cdot]_A: A \times A \rightarrow A$ satisfying the Jacobi identity, and a map $\rho: A \rightarrow TM$ which is compatible with the bracket in the sense that
\beq
	[f\mX,g\mY]_A = fg[\mX,\mY]_A + f\rho(\mX)(g)\mY - g\rho(\mY)(f)\mX, \;\; \forall \mX,\mY \in \Gamma(A), f,g \in C^{\infty}(M).
\eeq

An Atiyah Lie algebroid moreover possesses the following pair of short exact sequences:
\begin{equation} \label{Short Exact Sequence}
\begin{tikzcd}
0
\arrow{r} 
& 
L
\arrow{r}{j} 
\arrow[bend left]{l} 
& 
A
\arrow{r}{\rho} 
\arrow[bend left]{l}{\omega}
& 
TM
\arrow{r} 
\arrow[bend left]{l}{\sigma}
&
0\,.
\arrow[bend left]{l} 
\end{tikzcd}
\end{equation}
The upper short exact sequence is canonical and identifies the vertical subbundle of $A$, $V \equiv \text{im}(j) = \text{ker}(\rho)$ as isomorphic to the isotropy bundle $L = P \times_{\gravgroup} \mathfrak{g}_{\gravgroup} \oplus P \times_{\gaugegroup} \mathfrak{g}_{\gaugegroup} \equiv \gravL \oplus \gaugeL$. Here, $L$ may be thought of as a bundle of Lie algebras with standard fiber equivalent to the Lie algebra of the structure group $G$. In other words, we can think of elements in $L$ as consisting of direct sums of local generators of elements of $\gravgroup$ and $\gaugegroup$: $\un{\mu} = \un{\mu}_L \oplus \un{\mu}_G$. The lower short exact sequence defines a connection on $A$, $H \equiv \text{im}(\sigma) = \text{ker}(\omega)$, a horizontal subbundle on $A$ which is complimentary to $V$ such that $A = H \oplus V$ globally. 

The canonical maps $j: L \rightarrow A$ and $\rho: A \rightarrow TM$ are bracket morphisms:
\beq \label{canonical maps are morphisms}
	R^{j}(\un{\mu},\un{\nu}) = [j(\un{\mu}),j(\un{\nu})]_A - j([\un{\mu},\un{\nu}]_L) = 0, \qquad R^{\rho}(\mX,\mY) = [\rho(\mX),\rho(\mY)]_{TM} - \rho([\mX,\mY]_A) = 0.
\eeq
In \eqref{canonical maps are morphisms} we have introduced the notation
\beq
	R^{f}(a,b) = [f(a),f(b)]_{X} - f([a,b]_Y)
\eeq
to denote the curvature of the map $f: Y \rightarrow X$ where $X$ and $Y$ are spaces with brackets $[\cdot,\cdot]_X$ and $[\cdot,\cdot]_Y$, respectively. Conversely, the connection maps $\omega: A \rightarrow L$ and $\sigma: TM \rightarrow A$ are not morphisms in general. Rather, the curvature of $\omega$ and $\sigma$ quantify the failure of the horizontal distribution to close under the bracket on $A$. A generic element on $A$ can be expressed globally as $\mX = \sigma(\un{X}) \oplus j(\un{\mu})$ where $\un{X} \in TM$ and $\un{\mu} \in L$. Thus, we can define the bracket on $A$ directly from the brackets on $TM$ and $L$ along with the curvature of the structure maps:
\begin{multline} \label{Bracket on A}
	[\sigma(\un{X}) \oplus j(\un{\mu}), \sigma(\un{Y}) \oplus j(\un{\nu})]_A = \; \sigma\bigg([\un{X},\un{Y}]_{TM}\bigg) \\
	\oplus j\bigg([\un{\mu},\un{\nu}]_L - R^{-\omega}(\sigma(\un{X}),\sigma(\un{Y})) - R^{-\omega}(\sigma(\un{X}),j(\un{\nu})) + R^{-\omega}(\sigma(\un{Y}),j(\un{\mu})) \bigg) 
\end{multline}
Notice that, in the case that $-\omega$ is a morphism, the bracket on $A$ reduces to the bracket on $TM \oplus L$. When this is not the case there are two additional contributions -- the first from $R^{-\omega}(\sigma(\un{X}),\sigma(\un{Y}))$ which quantifies the non-integrability of $H$, and the second from $R^{-\omega}(\sigma(\un{X}),j(\un{\mu}))$ which quantifies the fact that the generators in $L$ are sections of a vector bundle over $M$.\footnote{In physical terminology, they are spatially dependent.}

The map\footnote{Here, $\Omega^{\bullet}(A;E)$ denotes the exterior algebra of $A$ with values in the Lie algebroid representation $E$. See \cite{Ciambelli:2021ujl, Jia:2023tki, Klinger:2023qna} for an overview of this topic.} $\omega \in \Omega^1(A;L)$ is called the \emph{connection reform} and subsumes the more familiar role played by gauge fields and the spin connection in traditional approaches to gauge theory. To be precise the connection reform splits as
\beq
	\omega = \omega_L \oplus \omega_G
\eeq
where $\omega_L \in \Omega^1(A;\gravL)$ corresponds to the spin connection, and $\omega_G \in \Omega^1(A;\gaugeL)$ corresponds to the internal gauge fields. The connection reform $\omega$ has curvature
\beq
	\Omega = \hatd \omega + \frac{1}{2}[\omega \wedge \omega]_L = \hatd \omega_{L} + \frac{1}{2}[\omega_L \wedge \omega_L]_{\gravL} \oplus \hatd \omega_G + \frac{1}{2}[\omega_G \wedge \omega_G]_{\gaugeL} \equiv \Omega_L \oplus \Omega_G,
\eeq
where $\Omega_L$ corresponds to the curvature of the spin connection, and $\Omega_G$ is related to the gauge field strength. The curvature forms are all horizontal: $\Omega \in \Omega^2(H;L)$. 

Since we are considering a structure group which features $\gravgroup \subset GL(d;\mathbb{R})$, we must also specify a solder form in order to complete the geometric framing of the theory. Recall that a solder form is a map $\solder: A \rightarrow \mathcal{E}$, where $\mathcal{E}$ is a $d$-dimensional vector bundle associated with $\gravgroup$. In the simple case where $\gravgroup = SO(1,d-1)$, the solder form serves to encode the $\gravL$ invariant inner product and is subsequently related to the spacetime metric. We will not restrict ourselves to this case, so for our purpose we view $\solder$ simply as a device for framing the algebroid $A$. Like the curvature form, the solder form is horizontal: $\solder \in \Omega^1(H;\mathcal{E})$. 

The pair $(\omega,\solder)$ specifies the geometric data of $A$, and, in a pure gauge theory, would constitute the complete set of dynamical fields. However, in general one expects to include a host of fields which transform in representations associated with the structure group $G$. From the perspective of $A$ such fields organize themselves inside the family of exterior algebras $\Omega^{\bullet}(A;E)$ where each $E$ is a vector bundle over $M$ which admits a Lie algebroid representation \cite{Ciambelli:2021ujl, Jia:2023tki, Klinger:2023qna}
\beq
	\Aconn{E}: A \rightarrow \text{Der}(E).
\eeq
Thus, in general, we regard the full set of fields for a gauge theory formulated on a Lie algebroid as schematically by the triple $(\omega,\solder,\psi)$ where $\psi$ is a stand in for all of the matter fields transforming in representations associated with the geometric data. 

\section{Rigged Hilbert Spaces} \label{app: rigging}

In this short appendix we remind the reader of some relevant details about rigged Hilbert spaces (RHS). RHS were introduced by Gelfand in \cite{gel2016generalized} as a mathematical formalization of Dirac's bra-ket notation for quantum mechanics. More broadly, RHS provide a formalism in which elements of the continuous spectrum of a self adjoint operator can be associated with generalized eigenvectors, as elements of the discrete spectrum are associated with genuine eigenvectors. In constraint quantization, as discussed in Section \ref{sec: RAQ}, we seek solutions to the equation
\beq \label{constraint problem 1}
	U(\un{\psi}) = \un{\psi},
\eeq
where $U = \text{exp}(Q)$ is a unitary operator on $H$ obtained by exponentiating the self adjoint, but possibly unbounded, operator $Q$. The connection to RHS therefore arises by reformulating \eqref{constraint problem 1} as
\beq \label{constraint problem 2}
	Q(\un{\psi}) = 0.
\eeq
Because $Q$ generically has a continuous spectrum, solutions to \eqref{constraint problem 2} are not always contained in the Hilbert space. Nevertheless, there are solutions to \eqref{constraint problem 2} if we pass to an appropriate RHS in which $\un{\psi}$ can be interpreted as a generalized eigenvector with eigenvalue zero.

To begin, let us define the RHS. Our discussion follows closely that of \cite{Celeghini:2019mlj}. Let $H$ be an infinite dimensional, separable Hilbert space. A RHS associated with $H$ is a triple\footnote{This is often referred to as a Gelfand triple.} $(\Phi,H,\Phi^*)$ characterized by the sequence
\beq
	\Phi \hookrightarrow H \hookrightarrow \Phi^*.
\eeq
Here, $\Phi \subset H$ is a dense subspace of $H$, and $\Phi^*$ is the set of all continuous complex-linear functionals on $\Phi$. As a subspace, the embedding of $\Phi$ in $H$ is obtained trivially, while the embedding of $H$ in $\Phi^*$ is obtained by recognizing that any $\un{\psi} \in H$ can be promoted to a complex-linear map on $\Phi$ using the pairing of vectors defined by the inner product $g$ on $H$. It is important to recognize that $\Phi^*$ contains more than just the metric dual of $H$, which will be central importance moving forward. Let $A: \Phi \rightarrow H$ be a linear map with adjoint $A^{\dagger}: H \rightarrow \Phi$. If
\beq
	A^{\dagger}(\Phi) \subset \Phi, \qquad A^{\dagger}\rvert_{\Phi} \text{ is continuous},
\eeq
then $A$ can be extended to a continuous linear operator on $\Phi^*$, $A^{*}: \Phi^* \rightarrow \Phi^*$, such that
\beq \label{Rigged Action}
	\bigg(A^*(f)\bigg)(\un{\psi}) = f\bigg(A^{\dagger}(\un{\psi})\bigg), \; \forall f \in \Phi^*, \un{\psi} \in \Phi. 
\eeq

We now have all of the tools we require to state our main result which is the so-called Gelfand-Maurin theorem \cite{gel2016generalized, Maurin:1966zz}. Let $A$ be any self adjoint operator on $H$ and denote by $\sigma(A) \subseteq \mathbb{R}_+$ its spectrum. Then, there exists a RHS $\Phi \hookrightarrow H \hookrightarrow \Phi^*$ such that the following are true:
\begin{enumerate}
	\item For almost all $\lambda \in \sigma(A)$, in the Lebesgue sense, there exists $f_{\lambda} \in \Phi^*$ solving the generalized eigenvalue equation
	\beq
		A^*(f_{\lambda}) = \lambda f_{\lambda}. 
	\eeq
	\item For any pair of vectors $\un{\psi}, \un{\psi}' \in H$ and any Borel-measureable function $h: \sigma(A) \rightarrow \mathbb{C}$,
	\beq
		g\bigg(\un{\psi}, h(A)(\un{\psi}')\bigg) = \int_{\sigma(A)} \mu_L(\lambda) \; h(\lambda) \overline{f_{\lambda}(\un{\psi})} f_{\lambda}(\un{\psi}').
	\eeq
	Here $\mu_L$ is the standard Lebesgue measure on $\mathbb{R}_+$. 
	\item There exists an isometry $V: \Phi \rightarrow L_1(\sigma(A))$ such that
	\beq
		\bigg(V(\un{\psi})\bigg)(\lambda) = f_{\lambda}(\un{\psi}), \;\; \bigg(VAV^{-1}(\un{\psi})\bigg)(\lambda) = \lambda f_{\lambda}(\un{\psi}), \; \forall \un{\psi} \in \Phi.
	\eeq
	Here $L_1(\sigma(A))$ is the Hilbert space of Borel-measureable functions.
\end{enumerate} 

To conclude this Appendix, we can shed some light on the very formal results discussed by considering the most quintessential application of RHS: the spectrum of the position operator in quantum mechanics. Let $H = L^2(\mathbb{R},dx)$, and consider the operator
\beq
	\bigg(A(\psi)\bigg)(x) = x \psi(x). 
\eeq
Such an operator is unbounded and thus is only well defined on the dense subset of $H$ given by
\beq
	D(A) \equiv \{\psi \in H \; | \; \int_{\mathbb{R}} dx \; x^2 |\psi(x)|^2 < \infty \}. 
\eeq
In standard quantum mechanics we are taught that the operator $A$ has eigenvectors denoted by $\ket{x}$ which are meant to be represented by the ``functions" $\ket{x} \sim \delta(x)$. However, delta functions are formally distributions or generalized functions and do not belong to the Hilbert space $H$. Nevertheless, they do make sense as \emph{functionals} of $H$. Let us initiate the following notation:
\beq \label{delta functional}
	\delta: \mathbb{R}_+ \rightarrow H^*, \; \delta_x(\psi) = \psi(x). 
\eeq
Of course, \eqref{delta functional} is what one would write in the usual notation as
\beq
	\bra{x}\ket{\psi} = \int_{\mathbb{R}} dy \; \delta(x-y) \psi(y) = \psi(x). 
\eeq

Eq. \eqref{delta functional} is really the crux of the RHS approach. In terms of it, we can now reproduce the three conclusions of the Gelfand-Maurin theorem:
\begin{enumerate}
	\item Using \eqref{Rigged Action}, we have
	\beq
		\bigg(A^*(\delta_x)\bigg)(\psi) = \delta_x\bigg(A(\psi)\bigg) = \bigg(A(\psi)\bigg)(x) = x\psi(x) = x \delta_x(\psi). 
	\eeq
	Or, in other words
	\beq
		A^*(\delta_x) = x \delta_x. 
	\eeq
	\item Given any pair of elements $\psi, \psi' \in H$ and any function $h: \mathbb{R}_+ \rightarrow \mathbb{C}$ we can show by direct computation that:
	\beq
		g\bigg(\psi, h(A)(\psi')\bigg) = \int_{\mathbb{R}} dx \; \overline{\psi(x)} h(x) \psi'(x) = \int_{\mathbb{R}_+} dx \; h(x) \overline{\delta_x(\psi)} \delta_x(\psi'). 
	\eeq
	\item Finally, the isometry $V: H \rightarrow L_1(\mathbb{R}_+)$ simply identifies each function $\psi \in H$ with the components of its wavefunction: 
	\beq
		\bigg(V(\psi)\bigg)(x) = \delta_x(\psi) = \psi(x). 
	\eeq
\end{enumerate}

\section{Evaluating the Faddeev-Popov Determinant} \label{app: FP}

In Section \ref{sec: Path} we defined the Faddeev-Popov determinant as
\beq
	\Delta_{F}(x)^{-1} \equiv \int_{G} \mu(g) \; \delta(F \circ a_g(x)).
\eeq
As always, we can evaluate \eqref{FP Det} by changing variables in the integral:
\beq \label{FP Det 2}
	\Delta_{F}(x)^{-1} = \int \mathcal{D}F \; \text{det}\bigg\rvert \frac{\delta g}{\delta F \circ a_g(x)} \bigg\rvert \delta(F) = \text{det}\bigg\rvert \frac{\delta g}{\delta F \circ a_g(x)} \bigg\rvert_{F = 0}.
\eeq 
Let us denote the (inverse) functional Jacobian by $D_F$ which can be further massaged into the form
\beq
	D_F(x) = \frac{\delta F}{\delta x} \frac{\delta a_g(x)}{\delta g}.
\eeq
Then,
\beq	
	\Delta_{F}(x) = \text{det}\bigg\rvert D_F(x) \bigg\rvert_{F = 0}.
\eeq
In this form, we can use basic properties of Grassmann valued fields to rewrite the Faddeev-Popov determinant as a Berezin integral
\beq
	\Delta_{F}(x) = \text{Ber}_{c,\overline{c}}\bigg(e^{iS_{G}(x,c,\overline{c})}\bigg). 
\eeq
Here $c \in \mathfrak{g}$ and $\overline{c} \in \mathfrak{g}^*$ are Grassmann valued scalars called ``ghosts". The ghost action is of the form
\beq
	S_{G}(x,c,\overline{c}) = \int \overline{c} D_F(x)(c). 
\eeq

\providecommand{\href}[2]{#2}\begingroup\raggedright\endgroup

\end{document}

%% file: packages.tex
\usepackage{amsmath,amssymb,amsfonts,bbm,bm}
\usepackage{slashed}

\usepackage{tocloft} 

\usepackage[nosort]{cite} 
\usepackage{graphicx,color}
\usepackage[colorlinks=true, linkcolor=blue, citecolor=blue, linktoc=all]{hyperref}

\usepackage[usenames,dvipsnames]{xcolor}

\usepackage{tikz-cd}
\usepackage{pict2e}
\usepackage{physics}
\usepackage{comment} 
\usepackage[makeroom]{cancel}
\usepackage{musicography}

%% file: standard.tex
\newcommand{\beq}{\begin{eqnarray}}
\newcommand{\eeq}{\end{eqnarray}}
\newcommand{\beqn}{\begin{eqnarray}}
\newcommand{\eeqn}{\end{eqnarray}}
\newcommand{\bea}{\begin{eqnarray}}
\newcommand{\eea}{\end{eqnarray}}
\newcommand{\be}{\begin{equation}}
\newcommand{\ee}{\end{equation}}

\newcommand{\ack}[1]{[{\bf Pfft!: {#1}}]}

\newcommand{\un}[1]{\underline{#1}}
\def\pa{\partial}
\usepackage[normalem]{ulem}
\renewcommand{\uuline}[1]{\underline{\underline{#1}}}
\newcommand{\RR}{\mathbb{R}}

\newcommand{\CC}{\mathbb{C}}

\newcommand{\uiuc}[1]{
	\centerline{
		\begin{minipage}[c]{0.7\textwidth}
			\begin{center}
			${}^{#1}$ Illinois Center for Advanced Studies of the Universe \& Department of Physics,\\ 
			University of Illinois, 1110 West Green St., Urbana IL 61801, U.S.A.
			\end{center}
		\end{minipage}
		}
	}

%% file: AlgMacros.tex
\usepackage{mathrsfs,mathtools}
\renewcommand\mathbb[1]{\mathbbm{#1}}
\newcommand\mathbbb[1]{\boldsymbol{\mathbb{#1}}}

\newcommand{\Lagr}{\mathfrak{L}} 

\newcommand{\hatd}{\hat{d}}

\newcommand{\hatcon}[1]{\hat{i}_{#1}}

\newcommand{\mX}{\un{\mathfrak{X}}}
\newcommand{\mY}{\un{\mathfrak{Y}}}

\newcommand{\Aconn}[1]{\phi_{#1}}

\newcommand{\morphalg}{\mathbbb{G}}
\newcommand{\pbalg}{\mathbbb{P}}

\newcommand{\fldspalg}{\mathbbb{F}}

\newcommand{\ellalg}{\mathbbb{L}}

\newcommand{\algcon}[1]{\hat{\text{I}}_{{#1}}}

\newcommand{\algcurrdens}[1]{\text{J}_{#1}}

\newcommand{\alginjo}{\mathbbb{j}}

\newcommand{\MCb}{\bm{\varpi}}

\newcommand{\ellsec}[1]{\bm{#1}}

\makeatletter
\DeclareRobustCommand{\loplus}{\mathbin{\mathpalette\dog@lsemi{+}}}
\DeclareRobustCommand{\lotimes}{\mathbin{\mathpalette\dog@lsemi{\times}}}
\DeclareRobustCommand{\roplus}{\mathbin{\mathpalette\dog@rsemi{+}}}
\DeclareRobustCommand{\rotimes}{\mathbin{\mathpalette\dog@rsemi{\times}}}

\newcommand{\dog@rsemi}[2]{\dog@semi{#1}{#2}{-90,90}}
\newcommand{\dog@lsemi}[2]{\dog@semi{#1}{#2}{270,90}}
\newcommand{\dog@semi}[3]{%
  \begingroup
  \sbox\z@{$\m@th#1#2$}%
  \setlength{\unitlength}{\dimexpr\ht\z@+\dp\z@\relax}%
  \makebox[\wd\z@]{\raisebox{-\dp\z@}{%
    \begin{picture}(1,1)
    \linethickness{\variable@rule{#1}}
    \roundcap
    \put(0.5,0.5){\makebox(0,0){\raisebox{\dp\z@}{$\m@th#1#2$}}}
    \put(0.5,0.5){\arc[#3]{0.5}}
    \end{picture}%
  }}%
  \endgroup
}
\newcommand{\variable@rule}[1]{%
  \fontdimen8  
  \ifx#1\displaystyle\textfont3\else
    \ifx#1\textstyle\textfont3\else
      \ifx#1\scriptstyle\scriptfont3\else
        \scriptscriptfont3\relax
  \fi\fi\fi
}
\DeclareRobustCommand{\loplus}{\mathbin{\mathpalette\dog@lsemi{+}}}